\documentclass[ALICE,manyauthors]{cernphprep}

\usepackage{graphicx}
\usepackage{amssymb}
\usepackage[loose]{units}
\usepackage{upgreek}
\usepackage{color}
\usepackage{hyperref}
\usepackage{lineno}
\usepackage{cite}

\newcommand{\pT}{\ensuremath{p_{\rm T}}}

\newcommand{\pp}{pp}
\newcommand{\PbPb}{Pb-Pb}
\newcommand{\s}{\ensuremath{\sqrt{s}}} 
\newcommand{\snn}{\sqrt{s_\mathrm{NN}}} 
\newcommand{\RAA}{\ensuremath{R_{\rm{AA}}}}
\newcommand{\TAA}{\ensuremath{T_{\rm{AA}}}}
\newcommand{\Npart}{\ensuremath{N_{\rm{part}}}}

\newlength{\figwidthnarrow}
\newlength{\figwidthwide}
\begin{document}%
%%%%%%%%%%%%% ptdr definitions %%%%%%%%%%%%%%%%%%%%%
%
%%%%%%%%%%%%%%%  Title page %%%%%%%%%%%%%%%%%%%%%%%%
%
\begin{titlepage}
\PHyear{2014}
\PHnumber{091}    % required, obtained from PH
\PHdate{13 May}              % required
%\EXPnumber{ALICE-INT-2010-9999}     % optional
%\EXPdate{12 October 2010}           % optional
%
%%% Put your own title + short title here:
\title{Neutral pion production at midrapidity in pp and \PbPb\ collisions at $\mathbf{\snn = \unit[2.76]{\mathbf{TeV}}}$}
\ShortTitle{$\pi^0$ production in pp and \PbPb\ collisions at 2.76 TeV}   % appears on right page headers
%
%%% Do not change the next lines!
\Collaboration{ALICE Collaboration%
         \thanks{See Appendix~\ref{app:collab} for the list of collaboration
                      members}}
\ShortAuthor{ALICE Collaboration}      % appears on left page headers, do not change
\begin{abstract}
Invariant yields of neutral pions at midrapidity in the transverse
momentum range $0.6 < \pT < \unit[12]{GeV}/c$ measured in \PbPb\
collisions at $\snn = \unit[2.76]{TeV}$ are presented for six
centrality classes. The \pp\ reference spectrum was measured in the
range $0.4 < \pT < \unit[10]{GeV}/c$ at the same center-of-mass
energy. The nuclear modification factor, $\RAA$, shows a suppression
of neutral pions in central \PbPb\ collisions by a factor of up to
about $8-10$ for $5 \lesssim \pT \lesssim \unit[7]{GeV}/c$. The
presented measurements are compared with results at lower
center-of-mass energies and with theoretical calculations.

%%% Local Variables: 
%%% mode: plain-tex
%%% TeX-master: "ALICE_PbPb2pi0"
%%% End: 

\end{abstract}
\end{titlepage}
%-----------------------------------------------------------------------------
\section{Introduction}
\label{Sect:Intro}

Quantum chromodynamics (QCD) predicts a transition from hadronic
matter to a state of deconfined quarks and gluons, i.e., to the
quark-gluon plasma (QGP), at a temperature of $T_c \approx
\unit[150-160]{MeV}$ at vanishing net baryon number
\cite{Borsanyi:2010cj,Bazavov:2011nk}. Energy densities created in
\PbPb\ collisions at the LHC are estimated to be sufficiently large
to reach this state \cite{Chatrchyan:2012mb,Alice:2011rta}. At low
transverse momenta (roughly $\pT \lesssim \unit[3]{GeV}/c$) it is
expected that pressure gradients in the QGP produced in an
ultrarelativistic collision of two nuclei give rise to a collective,
outward-directed velocity profile, resulting in a characteristic
modification of hadron spectra \cite{Heinz:2013th}. At sufficiently
large $\pT$ ($\gtrsim \unit[3-8]{GeV}$/c), hadrons in pp and \PbPb\
collisions originate from hard scattering as products of jet
fragmentation. Hard-scattered quarks and gluons, produced in the
initial stage of the heavy-ion collision, must traverse the QGP that
is produced around them and lose energy in the process
through interactions with that medium. This phenomenon (``jet
quenching'') leads to a modification of hadron yields at high \pT\
\cite{Bjorken:1982tu,Wang:1991xy}. By studying observables related to
jet quenching one would like to better understand the mechanism of
parton energy loss and to use hard probes as a tool to characterize
the QGP.

The modification of the hadron yields for different $\pT$ intervals 
in heavy-ion (A-A) collisions with respect to \pp\ collisions can be 
quantified with the nuclear modification factor
\begin{equation}
\RAA(\pT)=\frac{\mathrm{d}^2N/\mathrm{d}\pT \mathrm{d}y|_\mathrm{AA}}
{\langle \TAA \rangle \times \mathrm{d}^2\sigma/\mathrm{d}\pT \mathrm{d}y|_\mathrm{pp}}
\label{eq:raa}
\end{equation}
where the nuclear overlap function $\langle \TAA \rangle$ is related
to the average number of inelastic nucleon-nucleon collisions as
$\langle \TAA \rangle = \langle N_\mathrm{coll} \rangle /
\sigma_\mathrm{inel}^\mathrm{pp}$. In the factorization approach of a
perturbative QCD calculation of particle production from hard
scattering, the overlap function $\TAA$ can be interpreted as the
increase of the parton flux in going from pp to A-A
collisions. Without nuclear effects, \RAA\ will be unity in the hard
scattering regime.

Parton energy loss depends on a number of factors including the
transport properties of the medium and its space-time evolution, the
initial parton energy, and the parton type
\cite{Wiedemann:2009sh,d'Enterria:2009am,Majumder:2010qh,Armesto:2011ht,Burke:2013yra}.
The nuclear modification factor, \RAA, is also affected by the slope
of the initial parton transverse momentum spectrum prior to any
interaction with the medium and by initial-state effects like the
modifications of the parton distributions in nuclei. An important
constraint for modeling these effects comes from the study of p-A
collisions \cite{ALICE:2012mj}, but also from the study of A-A
collisions at different center-of-mass energies ($\snn$) and different
centralities. For instance, the increase in $\snn$ from RHIC to LHC
energies by about a factor 14 results in larger initial energy
densities and less steeply falling initial parton spectra
\cite{Horowitz:2011gd}. Moreover, at the LHC, pions with $\pT \lesssim
\unit[50]{GeV}/c$ are dominantly produced in the fragmentation of
gluons \cite{Sassot:2010bh}, whereas the contribution from quark
fragmentation in the same $\pT$ region is much larger and more
strongly varying with $\pT$ at RHIC \cite{Sassot:2009sh}. Therefore,
the pion suppression results at the LHC will be dominated by gluon
energy loss, and simpler to interpret than the results from
RHIC. Compared to measurements of the \RAA\ for inclusive charged
hadrons, differences between the baryon and meson \RAA\ provide
additional information on the parton energy loss mechanism and/or on
hadronization in A-A collisions
\cite{Sapeta:2007ad,Bellwied:2010pr}. Experimentally, neutral pions
are ideally suited for this as they can be cleanly identified (on a
statistical basis) via the decay $\pi^0 \rightarrow \gamma \gamma$.

The suppression of neutral pions and charged hadrons at large
transverse momentum
\cite{Adcox:2001jp,Adler:2002xw,Agakishiev:2011dc,Adare:2012wg,Adare:2013esx} 
and the disappearance of azimuthal
back-to-back correlations of charged hadrons in central Au-Au
collision at RHIC \cite{Adler:2002tq,Adams:2006yt} (see also
\cite{Arsene:2004fa,Adcox:2004mh,Back:2004je,Adams:2005dq}) were
interpreted in terms of parton energy loss in hot QCD matter. Neutral
pions in central Au-Au collisions at $\snn = \unit[200]{GeV}$ were
found to be suppressed by a factor of $4-5$ for $\pT \gtrsim
\unit[4]{GeV}/c$ \cite{Adler:2003qi,Adare:2008qa}. The rather weak
dependence of $\RAA$ on $\pT$ was described by a large number of jet
quenching models \cite{Bass:2008rv}. The $\snn$ and system size
dependence was studied in Cu-Cu collisions at $\snn = 19.4$, 62.4, and
\unit[200]{GeV} \cite{Adare:2008ad} and in Au-Au collisions at $\snn =
39$, 62.4, and \unit[200]{GeV} \cite{Adare:2012uk,Adare:2012wg}. In
central Cu-Cu collisions the onset of $\RAA < 1$ was found to occur
between $\snn = 19.4$ and \unit[62.4]{GeV}. For unidentified charged
hadrons in central \PbPb\ collisions at the LHC, \RAA\ was found to
increase from $\RAA < 0.2$ at $\pT \approx \unit[7]{GeV}/c$ to $\RAA
\approx 0.5$ for $\pT \gtrsim \unit[50]{GeV}/c$, in line with a
decrease of the relative energy loss with increasing parton \pT\
\cite{Aamodt:2010jd,CMS:2012aa,Abelev:2012hxa}.

The dependence of the neutral pion \RAA\ on $\snn$ and \pT\ in Au-Au
collisions at RHIC energies for $2 \lesssim \pT \lesssim
\unit[7]{GeV}/c$ is not fully reproduced by jet quenching calculations
in the GLV framework which is based on perturbative QCD
\cite{Adare:2012uk,Sharma:2009hn,Neufeld:2010dz}. This may indicate
that, especially for this intermediate $\pT$ range, jet quenching
calculations do not yet fully capture the relevant physics processes.
With the large increase in $\snn$ the measurement of \RAA\ at the LHC
provides a large lever arm to further constrain parton energy loss
models. Phenomena affecting pion production in the \pT\ range
$0.6<\pT< \unit[12]{GeV}/c$ of this measurement include collective
radial flow at low \pT\ and parton energy loss at high \pT.  The data
are therefore well suited to test models aiming at a description of
particle production over the full transverse momentum range, including
the potentially complicated interplay between jets and the evolving
medium.

%%% Local Variables: 
%%% mode: latex
%%% TeX-master: "ALICE_PbPb2pi0"
%%% End: 

\section{Detector description}
\label{sec:Detector}

Neutral pions were reconstructed via the two-photon decay channel
$\pi^0 \rightarrow \gamma \gamma$ which has a branching ratio of
98.8\% \cite{Beringer:1900zz}. Two independent methods of photon
detection were employed: with the Photon Spectrometer (PHOS) which is
an electromagnetic calorimeter \cite{Dellacasa:1999kd}, and with
photon conversions measured in the central tracking system using the
Inner Tracking System (ITS) \cite{Aamodt:2010aa} and the Time
Projection Chamber (TPC) \cite{Alme:2010ke}. In the latter method,
referred to as Photon Conversion Method (PCM), conversions out to the
middle of the TPC were reconstructed (radial distance $R \approx
\unit[180]{cm}$). The material in this range amounts to $(11.4 \pm
0.5)$\% of a radiation length $X_0$ for $|\eta| < 0.9$ corresponding
to a plateau value of the photon conversion probability of $(8.6 \pm
0.4)\%$. The measurement of neutral pions with two independent methods
with different systematics and with momentum resolutions having
opposite dependence on momentum provides a valuable check of the
systematic uncertainties and facilitates the measurements of neutral
pions in a wide momentum range with small systematic uncertainty.

PHOS consists of three modules installed at a distance of
\unit[4.6]{m} from the interaction point.  PHOS subtends
$260^\circ<\varphi<320^\circ$ in azimuth and $|\eta|<0.13$ in
pseudorapidity. Each module has 3584 detection channels in a matrix of
$64\times56$ cells made of lead tungstate (PbWO$_4$) crystals each of
size \unit[$2.2\times 2.2 \times 18$]{cm$^3$}. The transverse
dimensions of the cells are slightly larger than the PbWO$_4$
Moli\`{e}re radius of \unit[2]{cm}.  The signals from the cells are
measured by avalanche photodiodes with a low-noise charge-sensitive
preamplifier. In order to increase the light yield and thus to improve
energy resolution, PHOS crystals are cooled down to a temperature of
$-25~^\circ$C. The PHOS cells were calibrated in pp collisions by
equalizing the $\pi^0$ peak position for all cell combinations
registering a hit by a decay photon.

The Inner Tracking System (ITS) \cite{Aamodt:2008zz} consists of two
layers of Silicon Pixel Detectors (SPD) positioned at a radial
distance of \unit[3.9]{cm} and \unit[7.6]{cm}, two layers of Silicon
Drift Detectors (SDD) at \unit[15.0]{cm} and \unit[23.9]{cm}, and two
layers of Silicon Strip Detectors (SSD) at \unit[38.0]{cm} and
\unit[43.0]{cm}. The two SPD layers cover a pseudorapidity range of
$|\eta|<2$ and $|\eta| < 1.4$, respectively. The SDD and the SSD
subtend $|\eta|<0.9$ and $|\eta|<1.0$, respectively.

The Time Projection Chamber (TPC) \cite{Alme:2010ke} is a large
(85~m$^3$) cylindrical drift detector filled with a Ne/CO$_2$/N$_2$
(85.7/9.5/4.8\%) gas mixture. It covers a pseudorapidity range of
$|\eta|<0.9$ over the full azimuthal angle for the maximum track
length of 159 reconstructed space points. With the magnetic field of
$B=\unit[0.5]{T}$, electron and positron tracks were reconstructed
down to transverse momenta of about $\unit[50]{MeV}/c$. In addition,
the TPC provides particle identification via the measurement of the
specific energy loss (d$E$/d$x$) with a resolution of 5.5\%
\cite{Alme:2010ke}. The ITS and the TPC were aligned with respect to
each other to a precision better than $\unit[100]{\upmu m}$ using
tracks from cosmic rays and proton-proton collisions
\cite{Aamodt:2010aa}.

Two forward scintillator hodoscopes (VZERO-A and VZERO-C)
\cite{Cortese:2004aa} subtending $2.8 < \eta < 5.1$ and $-3.7 < \eta <
-1.7$, respectively, were used in the minimum bias trigger in the pp
and in the \PbPb\ run. The sum of the amplitudes of VZERO-A and
VZERO-C served as a measure of centrality in \PbPb\ collisions
\cite{Abelev:2013qoq}. Spectator (non-interacting) protons and
neutrons were measured with Zero Degree Calorimeters (ZDCs), located
close to the beam pipe, \unit[114]{m} away from the interaction point
on either side of the ALICE detector \cite{Aamodt:2008zz}.

%%% Local Variables: 
%%% mode: latex
%%% TeX-master: "ALICE_PbPb2pi0"
%%% End: 

\section{Data processing}
\label{sec:DataProcessing}

\subsection{Event selection}
\label{ssec:EventSelection}

The pp sample at $\sqrt{s}=2.76$~TeV was collected in the 2011 LHC
run. The minimum bias trigger (MB$_\mathrm{OR}$) in the pp run
required a hit in either VZERO hodoscope or a hit in the SPD. Based on
a van der Meer scan the cross section for inelastic pp collisions was
determined to be $\sigma_\mathrm{inel} = \unit[(62.8^{+2.4}_{-4.0} \pm
1.2)]{mb}$ and the MB$_\mathrm{OR}$ trigger had an efficiency of
$\sigma_\mathrm{MB_{OR}}/\sigma_\mathrm{inel} =
0.881^{+0.059}_{-0.035}$ \cite{Abelev:2012sea}. The results were
obtained from samples of $34.7 \times 10^6$ (PHOS) and $58 \times
10^6$ (PCM) minimum bias pp collisions corresponding to an integrated
luminosity ${\cal L}_{\rm int} = 0.63~\mbox{nb}^{-1}$ and ${\cal
  L}_{\rm int} = 1.05~\mbox{nb}^{-1}$, respectively. PHOS and the
central tracking detectors used in the PCM were in different readout
partitions of the ALICE experiment which resulted in the different
integrated luminosities.

The \PbPb\ data at $\snn = \unit[2.76]{TeV}$ were recorded in the 2010
LHC run. At the ALICE interaction region up to 114 bunches, each
containing about $7 \times 10^7$ $^{208}$Pb ions, were collided.  The
rate of hadronic interactions was about 100 Hz, corresponding to a
luminosity of about $1.3 \times \unit[10^{25}]{cm^{-2} s^{-1}}$. The
detector readout was triggered by the LHC bunch-crossing signal and a
minimum bias interaction trigger based on trigger signals from
VZERO-A, VZERO-C, and SPD \cite{Abelev:2013qoq}. The efficiency for
triggering on a hadronic \PbPb\ collision ranged between 98.4\% and
99.7\%, depending on the minimum bias trigger configuration. For the
centrality range 0-80\% studied in the \PbPb\ analyses $16.1 \times
10^6$ events in the PHOS analysis and $13.2 \times 10^6$ events in the
PCM analysis passed the offline event selection.

In both pp and \PbPb\ analyses, the event selection was based on VZERO
timing information and on the correlation between TPC tracks and hits
in the SPD to reject background events coming from parasitic beam
interactions. In addition, an energy deposit in the ZDCs of at least
three standard deviations above the single-neutron peak was required
for \PbPb\ collisions to further suppress electromagnetic interactions
\cite{Abelev:2013qoq}. Only events with a reconstructed vertex in $|
z_\mathrm{vtx} | < \unit[10]{cm}$ with respect to the nominal
interaction vertex position along the beam direction were used.

%%% Local Variables: 
%%% mode: latex
%%% TeX-master: "ALICE_PbPb2pi0"
%%% End: 

\subsection{Neutral pion reconstruction}
\label{ssec:pi0Reconstruction}

\begin{figure}[tb]
  \centering
  \includegraphics[width=\figwidthwide]{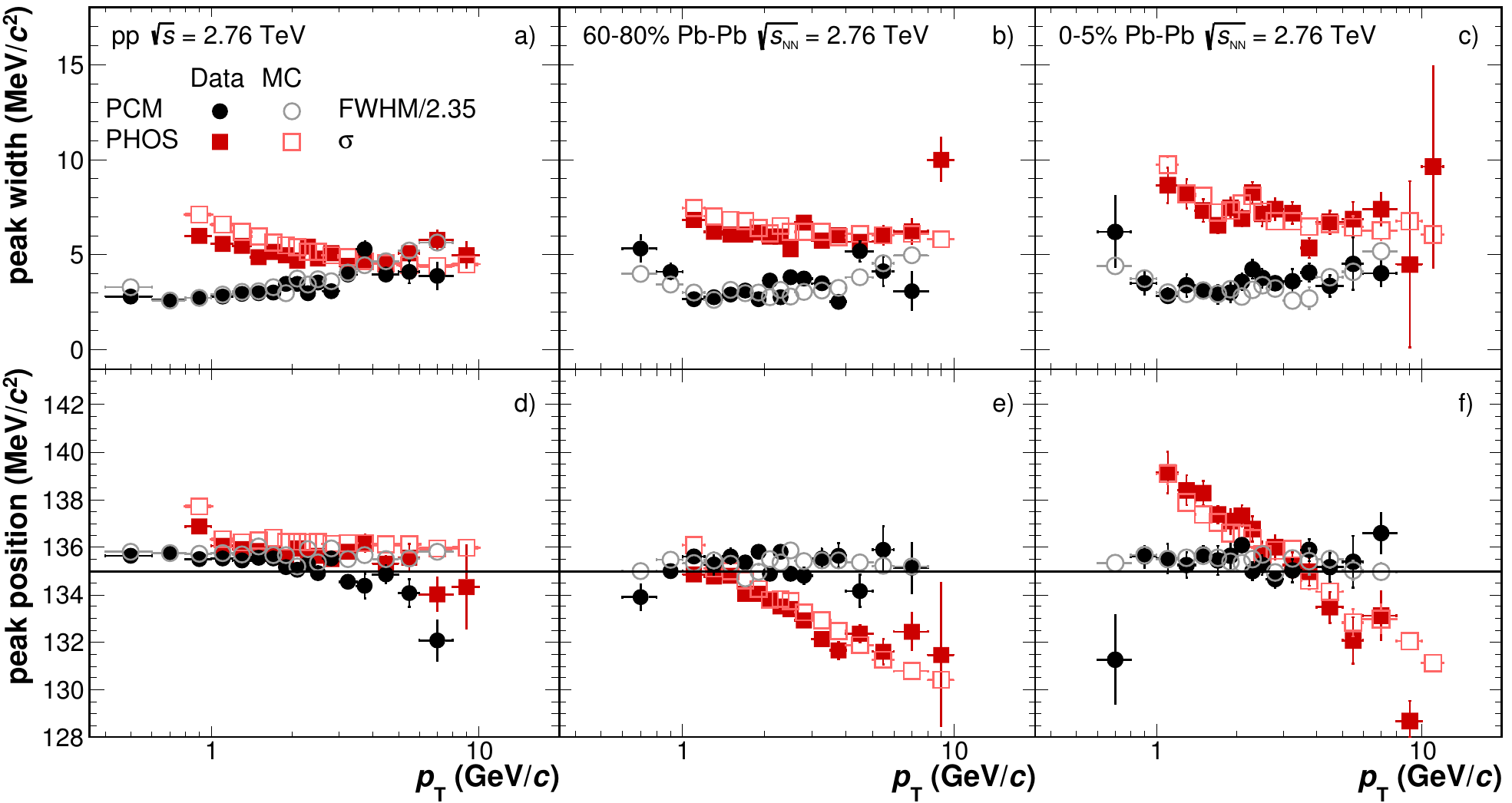}
  \caption{(Color online) Reconstructed $\pi^0$ peak width (upper row) and position (lower row)
    as a function
    of $\pT$ in \pp\ collisions at $\sqrt{s}=\unit[2.76]{TeV}$ (a, d),
    peripheral (b, e) and central (c, f) 
    \PbPb\ collisions at $\snn = \unit[2.76]{TeV}$ 
    in PHOS and in the photon
    conversion method (PCM) compared to Monte Carlo (MC) simulations. 
    The horizontal line in (d, e, f) indicates the nominal $\pi^0$
    mass.}
  \label{fig:InvMass}
\end{figure}

The PHOS and PCM analyses presented here are based on methods
previously used in pp collisions at $\sqrt{s} = 0.9$ and
$\unit[7]{TeV}$ \cite{Abelev:2012cn}. Neutral pions were reconstructed
using the $\pi^0\to \gamma\gamma$ decay channel either with both
photon candidates detected in PHOS or both photons converted into
$e^+e^-$ pairs and reconstructed in the central tracking system.
\begin{figure}[tb]
  \centering
  \includegraphics[width=\textwidth]{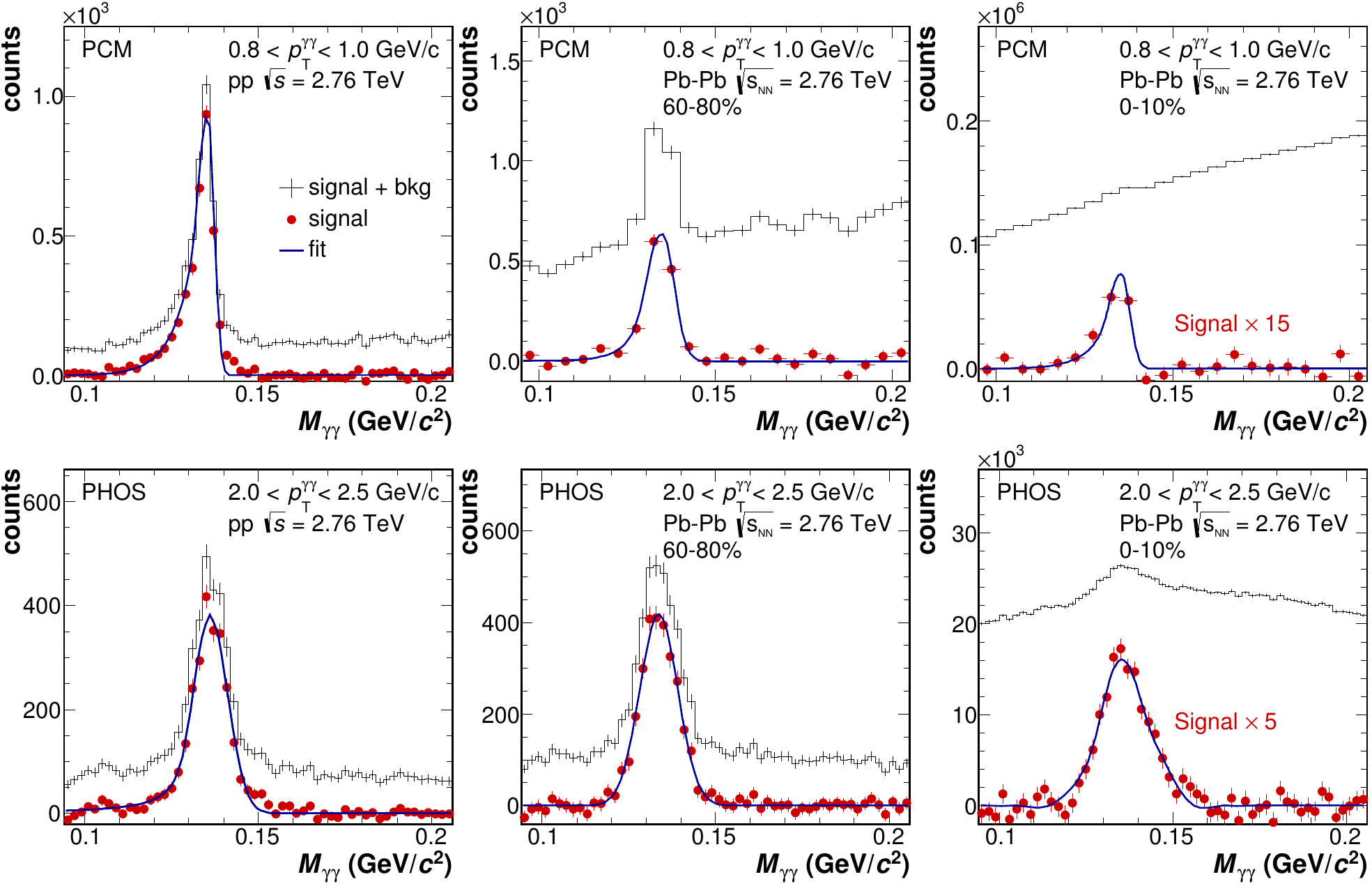}
  \caption{(Color online) Invariant mass spectra in selected \pT\
    slices for PCM (upper row) and PHOS (lower row) in the $\pi^0$
    mass region for \pp\ (left column), $60-80\%$ (middle column) and
    $0-10\%$ (right column) \PbPb\ collisions. The histogram and the
    filled points show the data before and after background
    subtraction, respectively. For the $0-10\%$ class the invariant
    mass distributions after background subtraction were scaled by a
    factor 15 and 5 for PCM and PHOS, respectively, for better
    visibility of the peak.  The positions and widths of the $\pi^0$
    peaks were determined from the fits, shown as blue curves, to the
    invariant mass spectra after background subtraction.}
  \label{fig:InvMassSpec}
\end{figure}
For the photon measurement with PHOS adjacent lead tungstate cells
with energy signals above a threshold ($\unit[12]{MeV}$) were grouped
into clusters \cite{Abelev:2014ffa}. The energies of the cells in a
cluster were summed up to determine the photon energy. The selection
of the photon candidates in PHOS was different for \pp\ and \PbPb\
collisions due to the large difference in detector occupancy. For pp
collisions cluster overlap is negligible and combinatorial background
small. Therefore, only relatively loose photon identification cuts on
the cluster parameters were used in order to maximize the $\pi^0$
reconstruction efficiency: the cluster energy for pp collisions was
required to be above the minimum ionizing energy
$E_\mathrm{cluster}>0.3$~GeV and the number of cells in a cluster was
required to be greater than two to reduce the contribution of hadronic
clusters. In the case of the most central \PbPb\ collisions about 80
clusters are reconstructed in PHOS, resulting in an occupancy of up to
1/5 of the 10752 PHOS cells. This leads to a sizable probability of
cluster overlap and to a high combinatorial background in the
two-cluster invariant mass spectra. A local cluster maximum was
defined as a cell with a signal at least \unit[30]{MeV} higher than
the signal in each surrounding cell. A cluster with more than one
local maximum was unfolded to several contributing clusters
\cite{Abelev:2014ffa}. As the lateral width of showers resulting from
hadrons is typically larger than the one of photon showers,
non-photonic background was reduced by a $\pT$ dependent shower shape
cut. This cut is based on the eigenvalues $\lambda_0$, $\lambda_1$ of
the covariance matrix built from the cell coordinates and weights
$w_i=\max[0,w_0+\log(E_i/E_\mathrm{cluster})]$, $w_0=4.5$ where $E_i$
is the energy measured in cell $i$. In the Pb-Pb case only cells with
a distance to the cluster center of $R_\mathrm{disp}=\unit[4.5]{cm}$
were used in the dispersion calculation. A 2D $\pT$-dependent cut in
the $\lambda_0$-$\lambda_1$ plane was tuned to have an efficiency of
$\sim 0.95$ using pp data. In addition, clusters associated with a
charged particle were rejected by application of a cut on the minimum
distance from a PHOS cluster to the extrapolation of reconstructed
tracks to the PHOS surface \cite{Abelev:2014ffa}. This distance cut
depended on track momentum and was tuned by using real data to
minimize false rejection of photon clusters. The corresponding loss of
the $\pi^0$ yield was about 1\% in pp collisions (independent of
$\pT$). In \PbPb\ collisions the $\pi^0$ inefficiency due to the
charged particle rejection is about 1\% in peripheral and increases to
about 7\% in central \PbPb\ collisions. In addition, to reduce the
effect of cluster overlap, the cluster energy was taken as the {\it
  core energy} of the cluster, summing over cells with centers within
a radius $R_\mathrm{core}=\unit[3.5]{cm}$ of the cluster center of
gravity, rather than summing over all cells of the cluster. By using
the core energy, the centrality dependence of the width and position
of the $\pi^0$ peak is reduced, due to a reduction of overlap effects.
The use of the core energy leads to an additional non-linearity due to
energy leakage outside $R_\mathrm{core}$: the difference between full
and core energy is negligible at $E_\mathrm{cluster} \lesssim
\unit[1]{GeV}$ and reaches $\sim 4$\% at $E_\mathrm{cluster} \sim
\unit[10]{GeV}$. This non-linearity, however, is well reproduced in
the GEANT3 Monte Carlo simulations \cite{Brun:1987ma} of the PHOS
detector response (compare $\pT$ dependences of peak positions in data
and Monte Carlo in Fig.\ \ref{fig:InvMass}) and is corrected for in
the final spectra.

PHOS is sensitive to pile-up from multiple events that occur within
the $\unit[6]{\upmu s}$ readout interval of the PHOS front-end
electronics. The shortest time interval between two bunch crossings in
pp collisions was \unit[525]{ns}. To suppress photons produced in
other bunch crossings, a cut on arrival time $|t|<\unit[265]{ns}$ was
applied to reconstructed clusters which removed 16\% of the clusters.
In the \PbPb\ collisions, the shortest time interval between bunch
crossing was \unit[500]{ns}, but the interaction probability per bunch
crossing was much smaller than in pp collisions. To check for a
contribution from other bunch crossings to the measured spectra, a
timing cut was applied, and the pile-up contribution was found to be
negligible in all centrality classes. Therefore, a timing cut was not
applied in the final PHOS Pb-Pb analysis.

The starting point of the conversion analysis is a sample of photon
candidates corresponding to track pairs reconstructed by a secondary
vertex (V0) finding algorithm \cite{Alessandro:2006yt,Abelev:2014ffa}.
In this step, no constraints on the reconstructed invariant mass and
pointing of the momentum vector to the collision vertex were applied.
Both tracks of a V0 were required to contain reconstructed clusters
(i.e., space points) in the TPC. V0's were accepted as photon
candidates if the ratio of the number of reconstructed TPC clusters
over the number of findable clusters (taking into account track
length, spatial location, and momentum) was larger than 0.6 for both
tracks. In order to reject $K_s^0$, $\Lambda$, and $\bar{\Lambda}$
decays, electron selection and pion rejection cuts were applied. V0's
used as photon candidates were required to have tracks with a specific
energy loss in the TPC within a band of [$-3 \sigma$, $5 \sigma$]
around the average electron d$E$/d$x$, and of more than $3\sigma$
above the average pion d$E$/d$x$ (where the second condition was only
applied for tracks with measured momenta $p > \unit[0.4]{GeV}/c$).
Moreover, tracks with an associated signal in the TOF detector were
only accepted as photon candidates if they were consistent with the
electron hypothesis within a $\pm 5 \sigma$ band. A generic particle
decay model based on the Kalman filter method \cite{CBM2007} was
fitted to a reconstructed V0 assuming that the particle originated
from the primary vertex and had a mass $M_{V0}=0$. Remaining
contamination in the photon sample was reduced by cutting on the
$\chi^2$ of this fit. Furthermore, the transverse momentum $q_T = p_e
\sin \theta_{V0,e}$ \cite{podolanski1954iii} of the electron, $p_e$,
with respect to the V0 momentum was restricted to $q_T <
\unit[0.05]{GeV}/c$. As the photon is massless, the difference $\Delta
\theta = |\theta_{e^-} - \theta_{e^+}|$ of the polar angles of the
electron and the positron from a photon conversion is small and the
bending of the tracks in the magnetic field only results in a
difference $\Delta \varphi = |\varphi_{e^-} - \varphi_{e^+}|$ of the
azimuthal angles of the two momentum vectors. Therefore, remaining
random track combinations, reconstructed as a V0, were suppressed
further by a cut on the ratio of $\Delta \theta$ to the total opening
angle of the $e^+e^-$ pair calculated after propagating both the
electron and the positron $\unit[50]{cm}$ from the conversion point in
the radial direction. In order to reject $e^+e^-$ pairs from Dalitz
decays the distance between the nominal interaction point and the
reconstructed conversion point of a photon candidate had to be larger
than \unit[5]{cm} in radial direction. The maximum allowed radial
distance for reconstructed V0's was \unit[180]{cm}.

Pile-up of neutral pions coming from bunch crossings other than the
triggered one also has an effect on the PCM measurement. At the level
of reconstructed photons, this background is largest for photons for
which both the electron and the positron were reconstructed with the
TPC alone without tracking information from the ITS. These photons,
which typically converted at large radii $R$, constitute a significant
fraction of the total PCM photon sample, which is about $67\%$ in case
of the pp analysis. This sample is affected because the TPC drift
velocity of $\unit[2.7]{cm/\upmu s}$ corresponds to a drift distance
of $\unit[1.41]{cm}$ between two bunch crossings in the pp run which
is a relatively short distance compared to the width of $\sigma_z
\approx \unit[5]{cm}$ of the distribution of the primary vertex in the
$z$ direction. The distribution of the distance of closest approach in
the $z$ direction ($\mathrm{DCA}_z$) of the straight line defined by
the reconstructed photon momentum is wider for photons from bunch
crossings other than the triggered one. The $\mathrm{DCA}_z$
distribution of photons which had an invariant mass in the $\pi^0$
mass range along with a second photon was measured for each $\pT$
interval. Entries in the tails at large $\mathrm{DCA}_z$ were used to
determine the background distribution and to correct the neutral pion
yields for inter bunch pile-up. For the pp analysis, this was a
$5-7\%$ correction for $\pT \gtrsim \unit[2]{GeV}/c$ and a correction
of up to $15\%$ at lower $\pT$ ($\pT \approx \unit[1]{GeV}/c$). In the
\PbPb\ case the correction at low $\pT$ was about 10\%, and became
smaller for higher $\pT$ and for more central collisions. For the
$20-40\%$ centrality class and more central classes the pile-up
contribution was negligible and no pile-up correction was applied. In
the PCM as well as in the PHOS analysis, events for which two or more
pp or \PbPb\ interactions occurred in the same bunch crossing were
rejected based on the number of primary vertices reconstructed with
the SPD \cite{Abelev:2014ffa} which has an integration time of less
than \unit[200]{ns}.

In the PHOS as well as in the PCM analysis, the neutral pion yield was
extracted from a peak above a combinatorial background in the
two-photon invariant mass spectrum. Examples of invariant mass
spectra, in the $\pi^0$ mass region, are shown in
Fig.~\ref{fig:InvMassSpec} for selected \pT\ bins for \pp\ collisions,
and peripheral and central \PbPb\ collisions. The combinatorial
background was determined by mixing photon candidates from different
events. In the PCM measurement the combinatorial background was
reduced by cutting on the energy asymmetry $\alpha
= |E_{\gamma_1}-E_{\gamma_2}|/(E_{\gamma_1}+E_{\gamma_2})$, where
$\alpha < 0.65$ was required for the central classes ($0-5\%$,
$5-10\%$, $10-20\%$, $20-40\%$) and $\alpha < 0.8$ for the two
peripheral classes ($40-60\%$, $60-80\%$). In both analyses the
mixed-event background distributions were normalized to the right and
left sides of the $\pi^0$ peak. A residual correlated background was
taken into account using a linear or second order polynomial fit. The
$\pi^0$ peak parameters were obtained by fitting a function, Gaussian
or a Crystal Ball function \cite{CrystalBall:1980} in the PHOS case or
a Gaussian combined with an exponential low mass tail to account for
bremsstrahlung \cite{Koch:2011} in the PCM case, to the
background-subtracted invariant mass distribution, see
Fig.~\ref{fig:InvMassSpec}. The Crystal Ball function was used in the
PHOS analysis of \pp\ data. A Gaussian was used alternatively to
determine systematic uncertainties of the peak parameters. In the
\PbPb\ case with worse resolution and smaller signal/background
ratios, the difference between Crystal Ball and Gaussian fits
disappeared and only the latter were used in the PHOS analysis. In the
case of PHOS the number of reconstructed $\pi^0$'s was obtained in
each $\pT$ bin by integrating the background subtracted peak within 3
standard deviations around the mean value of the $\pi^0$ peak
position. In the PCM analysis, the integration window was chosen to be
asymmetric ($m_{\pi^0}-0.035$~GeV/$c^2$, $m_{\pi^0}+0.010$~GeV/$c^2$)
to take into account the left side tail of the $\pi^0$ peak due to
bremsstrahlung energy loss of electrons and positrons from photon
conversions. In both analyses the normalization and integration
windows were varied to estimate the related systematic uncertainties.
The peak positions and widths from the two analyses are compared to
GEANT3 Monte Carlo simulations in Fig.~\ref{fig:InvMass} as a function
of $\pT$. The input for the GEANT3 simulation came from the event
generators PYTHIA~8 \cite{Sjostrand:2007gs} and PHOJET
\cite{Engel:1995sb} in the case of pp collisions (with roughly equal
number of events) and from HIJING \cite{Gyulassy:1994ew} in the case
of Pb-Pb collisions. For the PCM analysis the full width at half
maximum (FWHM) divided by $2\sqrt{2\ln 2} \approx 2.35$ is shown. Note
the decrease of the measured peak position with $\pT$ in \PbPb\
collisions for PHOS. This is due to the use of the core energy instead
of the full cluster energy. At low $\pT$ in central Pb-Pb collisions,
shower overlaps can increase the cluster energy thereby resulting in
peak positions above the nominal $\pi^0$ mass. A good agreement in
peak position and width between data and simulation is observed in
both analyses. The remaining small deviations in the case of PHOS were
taken into account as a systematic uncertainty related to the global
energy scale.

\begin{figure}[t]
\centering
\includegraphics[width=\figwidthnarrow]{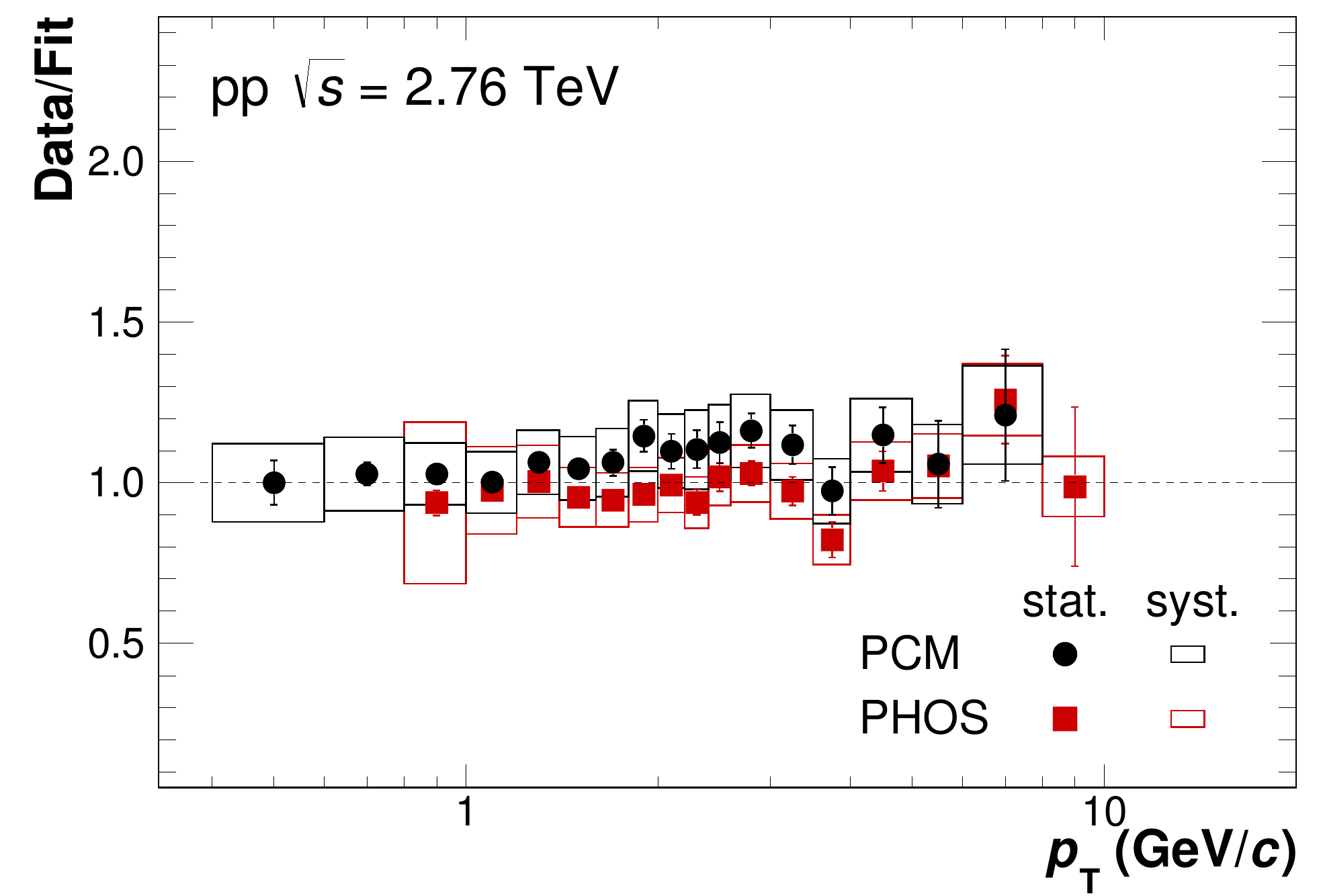}
\caption{(Color online) Ratio of the fully corrected $\pi^0$ spectra
  in pp collisions at $\sqrt{s}=\unit[2.76]{TeV}$ measured with PHOS and 
  PCM methods to the fit of the combined spectrum. Vertical lines 
  represent statistical uncertainties, the boxes systematic uncertainties.}
\label{fig:CompPPSpectra}
\end{figure}
The correction factor $\varepsilon(\pT)$ for the PHOS detector
response and the acceptance $A(\pT)$ were calculated with GEANT3 Monte
Carlo simulations tuned to reproduce the detector response.  The
factor $\varepsilon(\pT)$ takes the loss of neutral pions due to
analysis cuts, effects of the finite energy resolution and, in case of
Pb-Pb collisions, effects of shower overlaps into account. The shape
of the $\pi^0$ input spectrum needed for the calculation of
$\varepsilon(\pT)$ was determined iteratively by using a fit of the
corrected spectrum of a given pass as input to the next. In the case
of \PbPb\ collisions the embedding technique was used in the PHOS
analysis: the PHOS response to single $\pi^0$'s was simulated, the
simulated $\pi^0$ event was added to a real \PbPb\ event on the cell
signal level, after which the standard reconstruction procedure was
performed. The correction factor $\varepsilon(\pT) =
(N_\mathrm{rec}^\mathrm{after}(\pT)-N_\mathrm{rec}^\mathrm{before}(\pT))/N_\mathrm{sim}(\pT)$
was defined as the ratio of the difference of the number of
reconstructed $\pi^0$'s after and before the embedding to the number
of simulated $\pi^0$'s. In the \pp\ case, the PHOS occupancy was so
low that embedding was not needed and $\varepsilon(\pT)$ was obtained
from the $\pi^0$ simulations alone. Both in the \PbPb\ and the \pp\
analysis, an additional 2\% channel-by-channel decalibration was
introduced to the Monte Carlo simulations, as well as an energy
non-linearity observed in real data at low energies which is not
reproduced by the GEANT simulations. This non-linearity is equal to
$2.2\%$ at $\pT=1$~GeV/$c$ and decreases rapidly with \pT\ (less than
$0.5\%$ at $\pT>3$~GeV/$c$). For PHOS, the $\pi^0$ acceptance $A$ is
zero for $\pT<0.4$~GeV/$c$. The product $\varepsilon\cdot A$ increases
with $\pT$ and saturates at about $1.4\times 10^{-2}$ for a neutral
pion with $\pT>15$~GeV/$c$. At high transverse momenta
($\pT>25$~GeV/$c$) $\varepsilon$ decreases because of merging of
clusters of $\pi^0$ decay photons due to the decreasing average
opening angle of the $\pi^0$ decay photons. The correction factor
$\varepsilon$ does not show a centrality dependence for events in the
$20-80$\% class, but in the most central bin it increases by $\sim
10$\% due to an increase in cluster energies caused by cluster
overlap.

In the PCM, the photon conversion probability of about 8.6\% is
compensated by the large TPC acceptance. Neutral pions were
reconstructed in the rapidity interval $|y|<0.6$ and the decay photons
were required to satisfy $|\eta| < 0.65$. The $\pi^0$ efficiency
increases with $\pT$ below $\pT \approx \unit[4]{GeV}/c$ and remains
approximately constant for higher $\pT$ at values between $1.0 \times
10^{-3}$ in central collisions ($0-5\%$, energy asymmetry cut $\alpha
< 0.65$) and $1.5 \times 10^{-3}$ in peripheral collisions ($60-80\%$,
$\alpha < 0.8$). For the centrality classes $0-5\%$, $5-10\%$,
$10-20\%$, $20-40\%$, for which $\alpha < 0.65$ was used, the $\pi^0$
efficiency varies between $1.0 \times 10^{-3}$ and $1.2 \times
10^{-3}$. This small centrality dependence is dominated by the
centrality dependence of the V0 finding efficiency. Further
information on the PHOS and PCM efficiency corrections can be found in
\cite{Abelev:2014ffa}.
\begin{figure*}[tb]
\centering
\includegraphics[width=\figwidthwide]{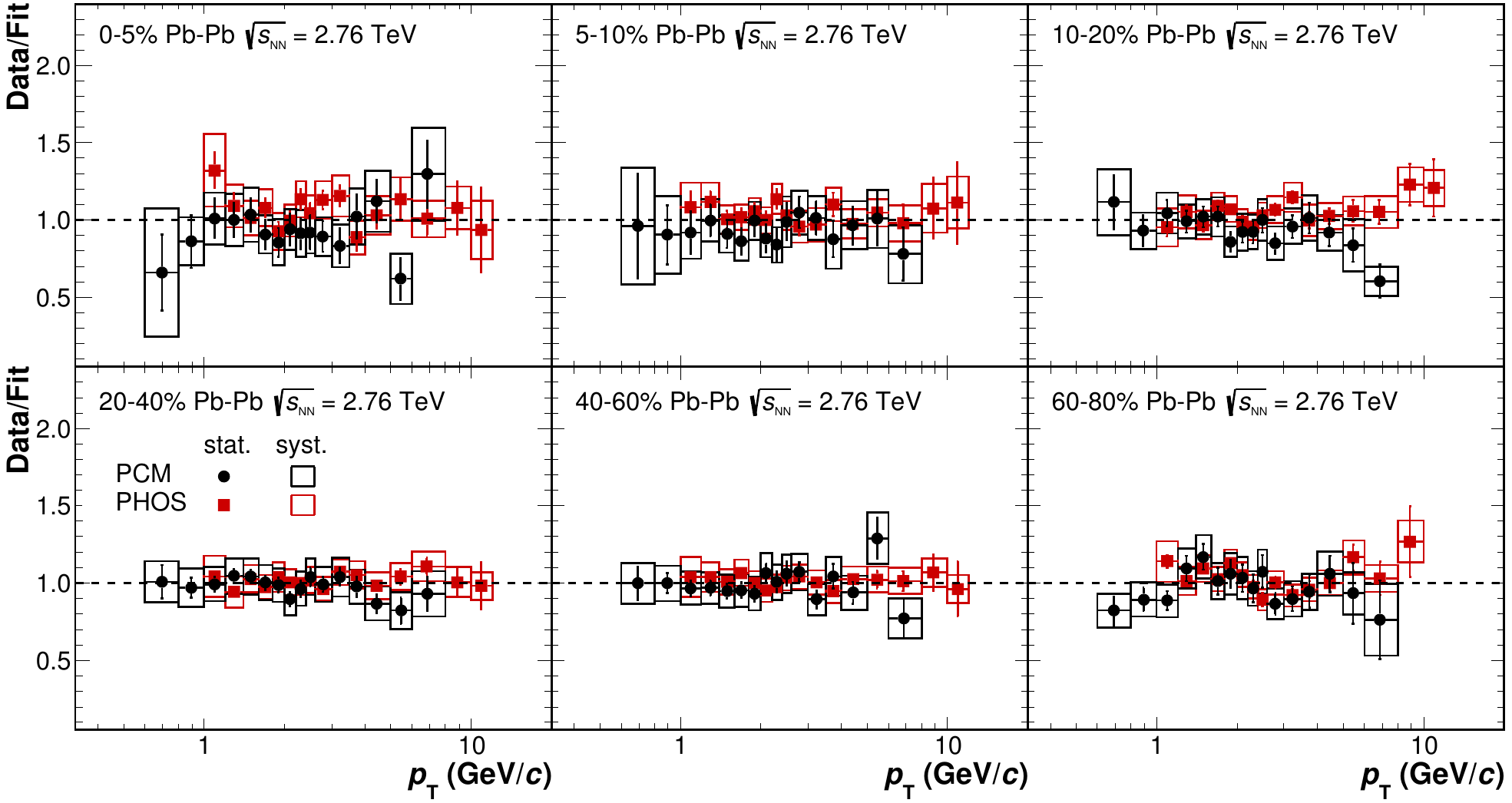}
\caption{(Color online) Ratio of the fully corrected $\pi^0$ spectra
  in Pb-Pb collisions at $\snn=\unit[2.76]{TeV}$ in six centrality
  bins measured with PHOS and PCM to the fits to the combined result
  in each bin. Vertical lines represent statistical uncertainties, the
  boxes the systematic uncertainties.}
\label{fig:CompPbPbSpectra}
\end{figure*}

The invariant differential neutral pion yield was calculated as
\begin{equation}
  E \frac{{\rm d}^3 N }{{\rm d}^3p} =
  \frac{1}{2\pi} \frac{1}{N_{\rm events}} \frac{1}{\pT} 
  \frac{1}{\varepsilon \,A}\frac{1}{\mathit{Br}}\frac{N^{\pi^0}}{\Delta y \Delta \pT},
\end{equation}
where $N_{\rm events}$ is the number of events; $\pT$ is the
transverse momentum within the bin to which the cross section has been
assigned after the correction for the finite bin width $\Delta \pT$,
$\mathit{Br}$ is the branching ratio of the decay $\pi^0\to
\gamma\gamma$, and $N^{\pi^0}$ is the number of reconstructed
$\pi^0$'s in a given $\Delta y$ and $\Delta \pT$ bin. Finally, the
invariant yields were corrected for the finite $\pT$ bin width
following the prescription in \cite{Lafferty:1994cj}, i.e., by
plotting the measured average yield at a $\pT$ position for which the
differential invariant yield coincides with the bin average. Secondary
$\pi^0$'s from weak decays or hadronic interactions in the detector
material were subtracted using Monte Carlo simulations. The
contribution of $\pi^0$'s from K$^0_\mathrm{s}$ as obtained from the
used event generators was scaled in order to reproduce the measured
K$^0_\mathrm{s}$ yields \cite{Abelev:2013xaa}. The correction for
secondary $\pi^0$'s was smaller than $2\%$ ($5\%$) for $\pT \gtrsim
\unit[2]{GeV}/c$ in the pp as well as in the \PbPb\ analysis for PCM
(PHOS).

\begin{table*}[tbh]
  \centering
  \resizebox{\linewidth }{!}{ %
  \begin{tabular}{|c|c|c|c|c|c|c|} 
   \hline
                          &  \multicolumn{6}{c|}{PHOS}   \\
   \cline{2-7}
                          &  \multicolumn{2}{c|}{\pp} &
                          \multicolumn{2}{c|}{\PbPb, $60-80\%$} &  \multicolumn{2}{c|}{\PbPb, $0-5\%$}  \\
   \cline{2-7}
                          & 1.1~GeV/$c$ &  7.5~GeV/$c$ & 3~GeV/$c$     &  10~GeV/$c$ &  3~GeV/$c$  & 10~GeV/$c$ \\
    \hline
    Yield extraction      & $ 8$     & $ 2.3$     &  $ 0.8$ & $ 6.8$ &  $ 3.7$    & $ 5.7$    \\
    Photon identification          & --    & --              &  $ 1.7$  & $ 1.7$  &   $ 4.4$    & $ 4.4$  \\ 
    Global $E$ scale     & $ 4$   & $ 6.2 $   &  $ 4.1$  & $ 5.3$ &  $ 6.1$    & $ 7.8$    \\
    Non-linearity         & $ 9$      & $ 1.5 $   &  $ 1.5$ & $ 1.5$ &  $ 1.5$    & $ 1.5$   \\
    Conversion            & $ 3.5$   & $ 3.5 $   &  $ 3.5$  &$ 3.5$ &  $ 3.5$    & $ 3.5$    \\
    Module alignment  & $ 4.1$   & $ 4.1 $   &  $ 4.1$  &$ 4.1$ &  $ 4.1$    & $ 4.1$  \\
    Other                     & $ 2$      & $ 1.4 $   &  $ 2.4$  & $ 2.4$ &  $ 3.1$    & $ 3.4$    \\ 
    \hline    
   Total                       & $ 13.9$    & $ 8.8$  & $7.6$& $10.7$ &  $10.7$   & $12.7$  \\
    \hline
%  \end{tabular}
%\vspace{1cm}
%  \begin{tabular}{|c|c|c|} 
   \hline
                          &  \multicolumn{6}{c|}{PCM}   \\
   \cline{2-7}
                          &  \multicolumn{2}{c|}{\pp} &\multicolumn{2}{c|}{\PbPb, $60-80\%$} &  \multicolumn{2}{c|}{\PbPb, $0-5\%$}   \\
   \cline{2-7}
                              & 1.1~GeV/$c$ &  5.0~GeV/$c$  & 1.1~GeV/$c$ &  5.0~GeV/$c$ & 1.1~GeV/$c$ &  5.0~GeV/$c$ \\
    \hline
    Material budget                              &  $ 9.0$   & $ 9.0$  &$ 9.0$    & $ 9.0$ &  $ 9.0$  & $ 9.0$    \\
    Yield extraction                               &  $ 0.6$  &  $ 2.6$ &  $ 3.3$    & $ 5.9$&  $ 10.6$  & $  5.0 $  \\
    $e^{+}/e^{-}$ identification               &  $ 0.7$  &  $ 1.4$  &  $ 2.9$    & $ 5.3$&  $ 9.0$    & $ 10.5$  \\
    Photon identification ($\chi^2 (\gamma)$)   &  $ 2.4$  &  $ 0.9$  &  $ 3.7$    & $ 4.6$ &  $ 4.0$    & $ 6.7$  \\
    $\pi^0$ reconstruction efficiency      &  $ 0.5$  &  $ 3.6$  &  $ 3.5$    & $ 4.1$ &  $ 6.7$    & $ 8.4$  \\
    Pile-up correction                                   &  $ 1.8$   & $ 1.8$   &  2.0               & 2.0 &  --        & --            \\
    \hline 
    Total                           &  $ 9.5$  &  $ 10.3$   &  $ 11.4$   & $ 13.6$& $ 18.3$    & $ 18.2$ \\
    \hline
  \end{tabular}}
  \caption{Summary of the relative systematic uncertainties in percent
    for selected $\pT$ bins for the PHOS and the PCM analyses. }
  \label{tab:SysErrs}
\end{table*} 

A summary of the systematic uncertainties for two representative $\pT$
values in \pp, peripheral and central \PbPb\ collisions is shown in
Table~\ref{tab:SysErrs}. In PHOS, one of the largest sources of the
systematic uncertainty both at low and high $\pT$ is the raw yield
extraction. It was estimated by varying the fitting range and the
assumption about the shape of the background under the peak. In
central collisions, major contributions to the systematic uncertainty
are due to the efficiency of photon identification and the global
energy scale. The former was evaluated by comparing
efficiency-corrected $\pi^0$ yields, calculated with different
identification criteria. The latter was estimated by varying the
global energy scale within the tolerance which would still allow to
reproduce the peak position in central and peripheral
collisions. The uncertainty related to the non-linearity of the PHOS
energy response was estimated by introducing different non-linearities
into the Monte Carlo simulations under the condition that the simulated $\pT$
dependence of the $\pi^0$ peak position and peak width was still
consistent with the data. The uncertainty of the PHOS measurement
coming from the uncertainty of the fraction of photons lost due to
conversion was estimated by comparing measurements without magnetic
field to the measurements with magnetic field. 

In the PCM measurement, the main sources of systematic uncertainties
include the knowledge of the material budget, raw yield extraction,
electron identification (PID), the additional photon identification
cuts, and $\pi^0$ reconstruction efficiency. The uncertainty related
to the pile-up correction is only relevant in pp and peripheral \PbPb\
collisions. The contribution from the raw $\pi^0$ yield extraction was
estimated by changing the normalization range, the integration window,
and the combinatorial background evaluation. Uncertainties related to
the electron and photon identification cuts, and to the photon
reconstruction efficiency were estimated by evaluating the stability
of the results for different cuts. The total systematic uncertainties
of the PCM and the PHOS results were calculated by adding the
individual contributions in quadrature.

The comparisons of the fully corrected $\pi^0$ spectra measured by
PHOS and PCM in \pp\ and \PbPb\ collisions are presented in
Figs.~\ref{fig:CompPPSpectra} and \ref{fig:CompPbPbSpectra},
respectively. For a better comparison the PCM and PHOS data points
were divided by a function which was fitted to the combined spectrum.
In all cases, agreement between the two measurements is found. The
PHOS and PCM spectra were combined by calculating the average yields
together with their statistical and systematic uncertainties by using
the inverse squares of the total uncertainties of the PHOS and PCM
measurements for a given $\pT$ bin as respective weights
\cite{Beringer:1900zz}.

%%% Local Variables: 
%%% mode: latex
%%% TeX-master: "ALICE_PbPb2pi0"
%%% End: 

\begin{figure}[t] \centering
  \includegraphics[width=\figwidthnarrow]{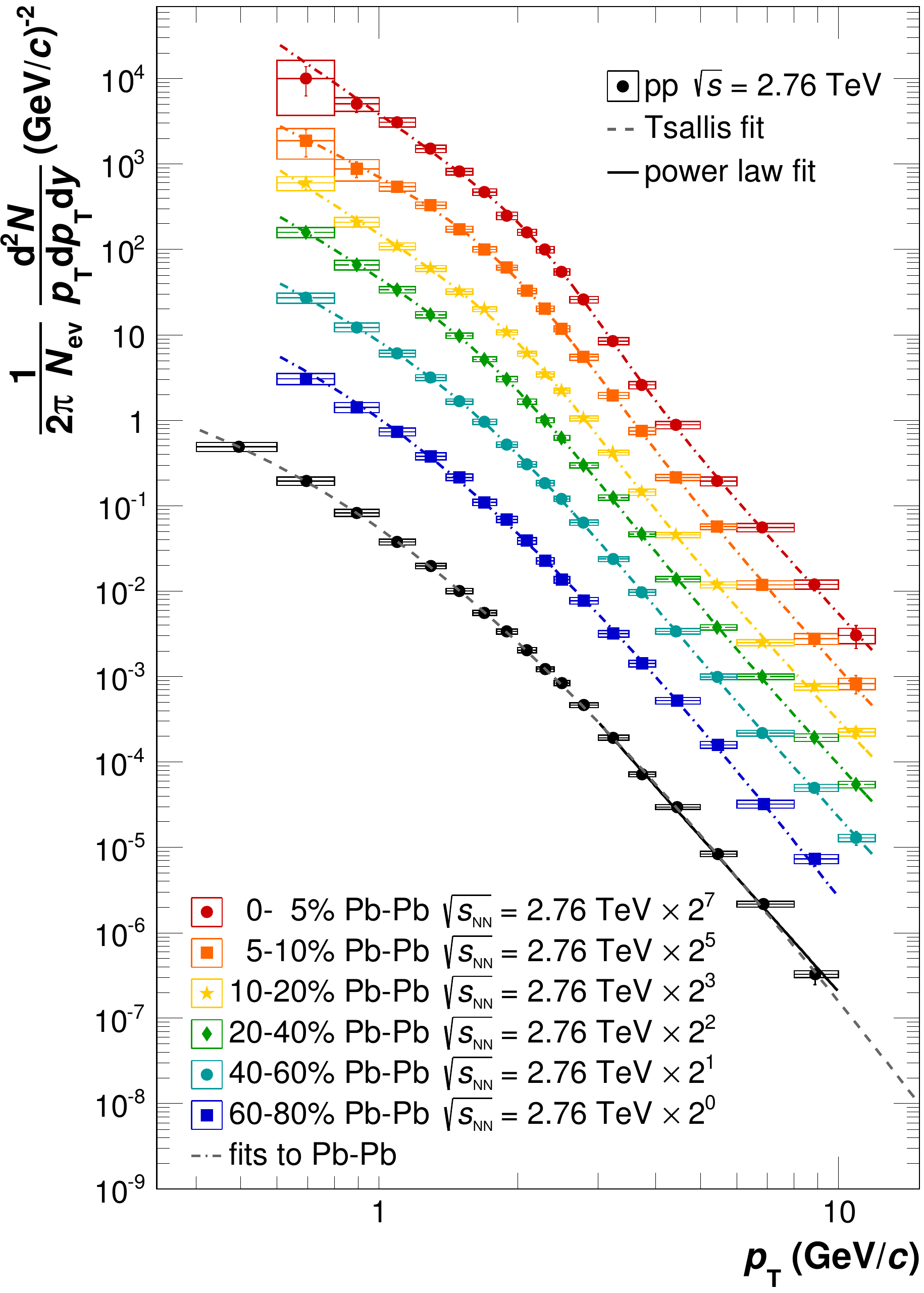}
  \caption{(Color online) Invariant differential yields of neutral
    pions produced in \PbPb\
    and inelastic \pp\ collisions at $\snn=\unit[2.76]{TeV}$. The
    spectra are the weighted average of the PHOS and the PCM results. The vertical
    lines show the statistical uncertainties, systematic uncertainties
    are shown as boxes. Horizontal lines indicate the bin width. The
    horizontal position of the data points within a bin was
    determined by the procedure described in \cite{Lafferty:1994cj}.
    For the \pp\ spectrum a fit with a power law function $1/\pT^n$
    for $\pT > \unit[3]{GeV}/c$ and a Tsallis function (also used in
    \cite{Abelev:2012cn}) are shown. The extrapolation of the pp
    spectrum provided by the Tsallis fit is used in the \RAA\
    calculation for $\pT \gtrsim \unit[8]{GeV}/c$.}
\label{fig:FinalSpectra}
\end{figure}

\section{Results}
\label{sec:Result}

The invariant neutral pion spectra measured in pp and \PbPb\
collisions are shown in Fig.~\ref{fig:FinalSpectra}. The \pT\ range
$0.6-\unit[12]{GeV}/c$ covered by the measurements includes the region
$\pT \approx \unit[7]{GeV}/c$ where the charged hadron \RAA\ exhibits
the strongest suppression
\cite{Aamodt:2010jd,CMS:2012aa,Abelev:2012hxa}. The invariant neutral
pion yield in inelastic \pp\ collisions shown in
Fig.~\ref{fig:FinalSpectra} is related to the invariant cross section
as $E\,\mathrm{d}^3\sigma/\mathrm{d}^3p =
E\,\mathrm{d}^3N/\mathrm{d}^3p \times \sigma_\mathrm{inel}$. Above
$\pT \approx \unit[3]{GeV}/c$ the pp spectrum is well described by a
power law $E\,\mathrm{d}^3N/\mathrm{d}^3p \propto 1/\pT^n$. A fit to
$\pT > \unit[3]{GeV}/c$ yields an exponent $n = 6.0 \pm 0.1$ with
$\chi^2/\mathrm{ndf} = 3.8/4$, which is significantly smaller than the
value $n = 8.22 \pm 0.09$ observed in pp collisions at $\sqrt{s} =
\unit[200]{GeV}$ \cite{Adare:2008qa}. 

\begin{figure}[htb]
  \centering
  \includegraphics[width=\figwidthnarrow]{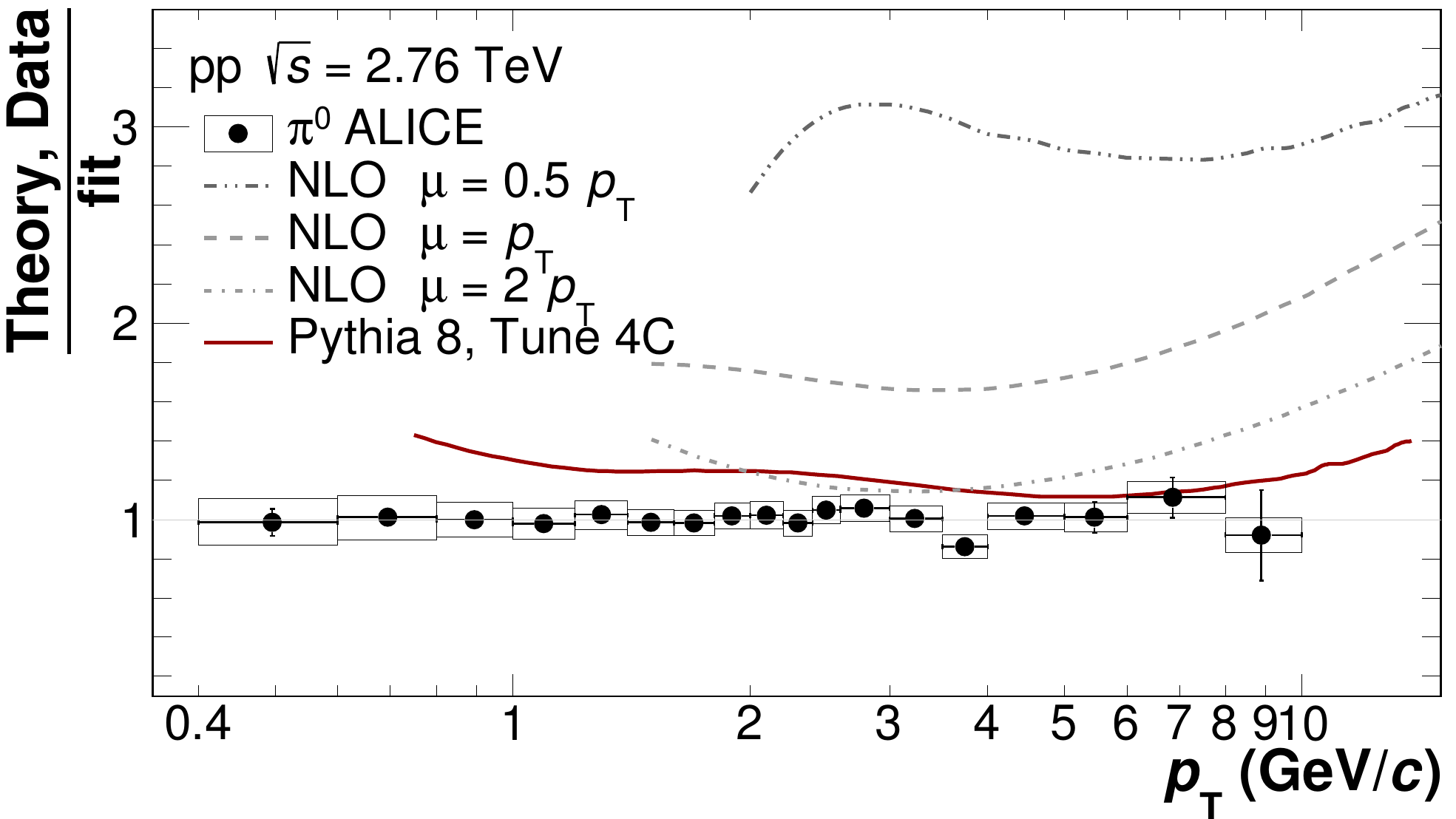}
  \caption{(Color online) Ratio of data or theory calculations to a
    fit of the neutral pion spectrum in pp collisions at
    $\snn = \unit[2.76]{TeV}$. The renormalization,
    factorization, and fragmentation scale of the next-to-leading
    order QCD calculation were varied simultaneously ($\mu = 0.5 \pT$,
    $\pT$, $2 \pT$). The calculation employed the CTEQ6M5
    \cite{Pumplin:2002vw} parton distribution functions and the DSS
    fragmentation function \cite{deFlorian:2007aj}. The solid red line is a comparison to the
    PYTHIA 8.176 (tune 4C) event generator \cite{Sjostrand:2007gs,Corke:2010yf}.}
\label{fig:cmp_theory_pp}
\end{figure}

Neutral pion production from hard scattering is dominated by the
fragmentation of gluon jets in the \pT\ range of the measurement. The
presented $\pi^0$ spectrum in pp collisions can therefore help
constrain the gluon-to-pion fragmentation function
\cite{d'Enterria:2013vba}. A next-to-leading-order (NLO) perturbative
QCD calculation employing the DSS fragmentation function
\cite{deFlorian:2007aj} agrees reasonably well with the measured
neutral pion spectrum at $\sqrt{s} = \unit[0.9]{TeV}$. At $\sqrt{s}
=\unit[7]{TeV}$, however, the predicted invariant cross sections are
larger than the measured ones \cite{Abelev:2012cn}. The comparison to
a NLO perturbative QCD calculation using the CTEQ6M5 parton
distributions \cite{Pumplin:2002vw} and the DSS fragmentation
functions in Fig.~\ref{fig:cmp_theory_pp} shows that the calculation
overpredicts the data already at $\sqrt{s}=\unit[2.76]{TeV}$ by a
similar factor as in pp collisions at $\sqrt{s}=\unit[7]{TeV}$. The
data are furthermore compared to a PYTHIA~8.176 (tune 4C)
\cite{Sjostrand:2007gs,Corke:2010yf} calculation which reproduces the
shape of the spectrum with an overall offset of about 20\%. It will be
interesting to see whether calculations in the framework of the color
glass condensate \cite{Lappi:2013zma}, which describe the neutral pion
spectrum in pp collisions at $\sqrt{s} = \unit[7]{TeV}$, will also
provide a good description of the data at $\sqrt{s} =
\unit[2.76]{TeV}$. 

The nuclear modification factor, \RAA, was calculated according to
Eq.~\ref{eq:raa}. For $\pT > \unit[8]{GeV}/c$ the extrapolation of the
pp spectrum provided by the power law fit shown in
Fig.~\ref{fig:FinalSpectra} was used as a reference. The systematic
uncertainty of the extrapolation was estimated based on the variation
of the fit range ($\pT>\unit[2, 3, 4]{GeV}/c$) and the systematic
uncertainty in the bin from $\pT = \unit[6-8]{GeV}/c$.  The average
values of the nuclear overlap function \TAA\ for each centrality class
were taken from \cite{Abelev:2013qoq} and are given in
Table~\ref{tab:Taa}. They were determined with a Glauber Monte Carlo
calculation \cite{Miller:2007ri,Alver:2008aq} by defining percentiles
with respect to the simulated impact parameter $b$ and therefore
represent purely geometric quantities.
\begin{table}[htb]
	\centering
	\begin{tabular}{|c|c|c|}
	\hline
	 centrality class & $\langle T_{AA} \rangle$ (1/mb)& rel. syst. uncert. (\%)\\ \hline
		$0-5\%$ & $26.32$ & $3.2$ \\
		$5-10\%$ & $20.56$ & $3.3$ \\
		$10-20\%$ & $14.39$ & $3.1$ \\
		$20-40\%$ & $6.85$ & $3.3$ \\
		$40-60\%$ & $1.996$ & $4.9$ \\
		$60-80\%$ & $0.4174$ & $6.2$ \\
	\hline
	\end{tabular}
 	\caption {Values for the overlap function $\langle T_{AA} \rangle$ for the centrality bins used in this analysis.}
        \label{tab:Taa}
\end{table}
\begin{figure}[ht]
  \centering
  \includegraphics[width=\figwidthnarrow]{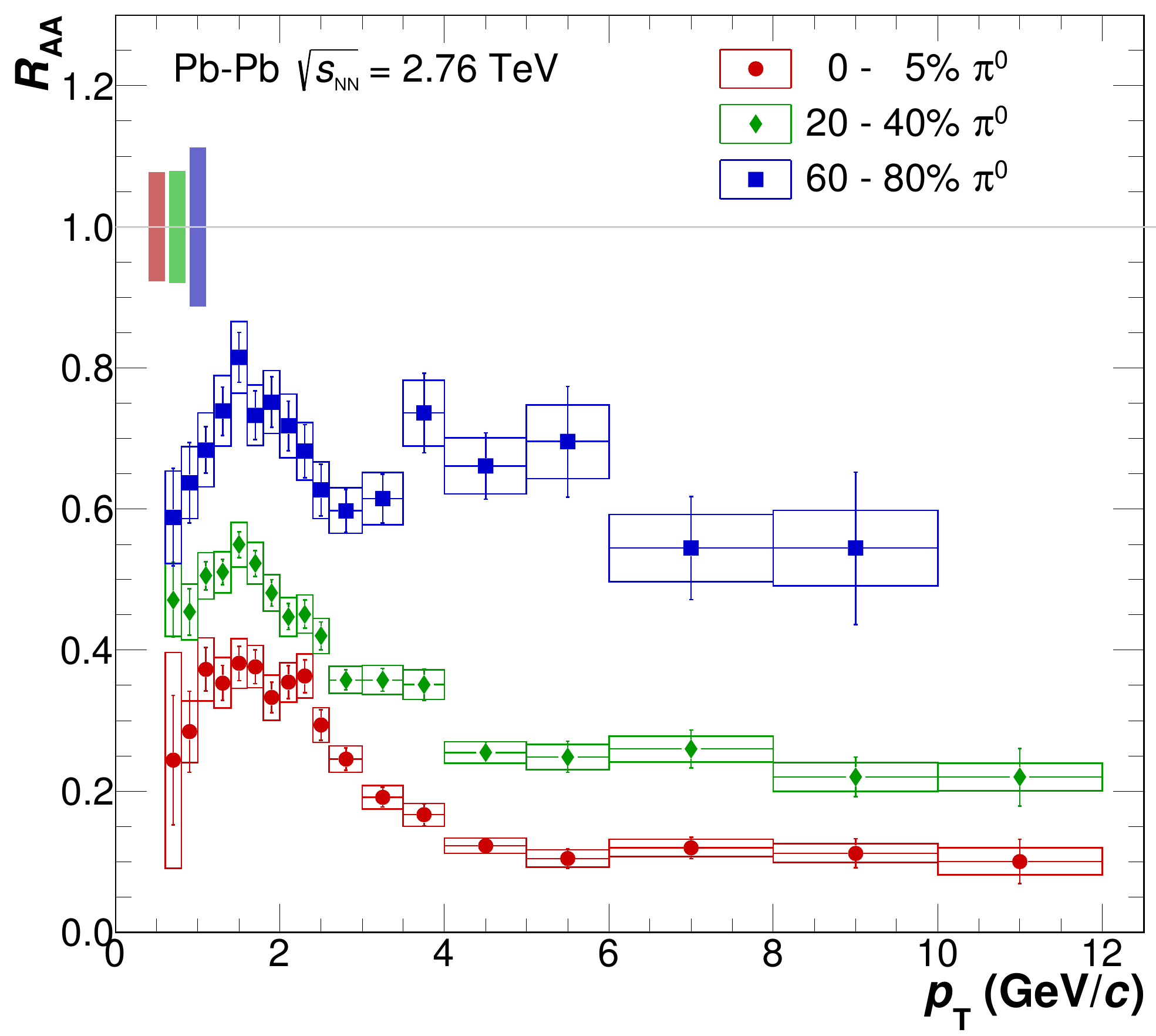}
  \caption{(Color online) Neutral pion nuclear modification factor
    \RAA\ for three different centralities ($0-5\%$, $20-40\%$,
    $60-80\%$) in \PbPb\ collisions at $\snn =
    \unit[2.76]{TeV}$. Vertical error bars reflect statistical
    uncertainties, boxes systematic uncertainties. Horizontal bars
    reflect the bin width. The boxes around unity reflect the
    uncertainty of the average nuclear overlap function ($T_{AA}$) and
    the normalization uncertainty of the pp spectrum added in
    quadrature. }
  \label{fig:RAA_All}
\end{figure}
The combined \RAA\ was calculated as a weighted average of the
individual \RAA\ measured with PHOS and PCM. This has the advantage of
reduced systematic uncertainties of the combined result. In
particular, the dominant uncertainty in the PCM, related to the
material budget, cancels this way. The results for the combined \RAA\
are shown in Fig.~\ref{fig:RAA_All}. In all centrality classes the
measured \RAA\ exhibits a maximum around $\pT \approx
\unit[1-2]{GeV}/c$, a decrease in the range $2 \lesssim \pT \lesssim
\unit[3-6]{GeV}/c$, and an approximately constant value in the
measured \pT\ range for higher \pT. For $\pT \gtrsim \unit[6]{GeV}/c$,
where particle production is expected to be dominated by fragmentation
of hard-scattered partons, \RAA\ decreases with centrality from about
$0.5-0.7$ in the $60-80$\% class to about 0.1 in the 0-5\% class. The
\RAA\ measurements for neutral pions and charged pions
\cite{Abelev:2014laa} agree with each other over the entire \pT\ range
for all centrality classes. Agreement between the neutral pion and
charged particle \RAA\ \cite{Abelev:2012hxa} is observed for $\pT
\gtrsim \unit[6]{GeV}/c$.

\begin{figure}[htb]
\centering
\includegraphics[width=\figwidthnarrow]{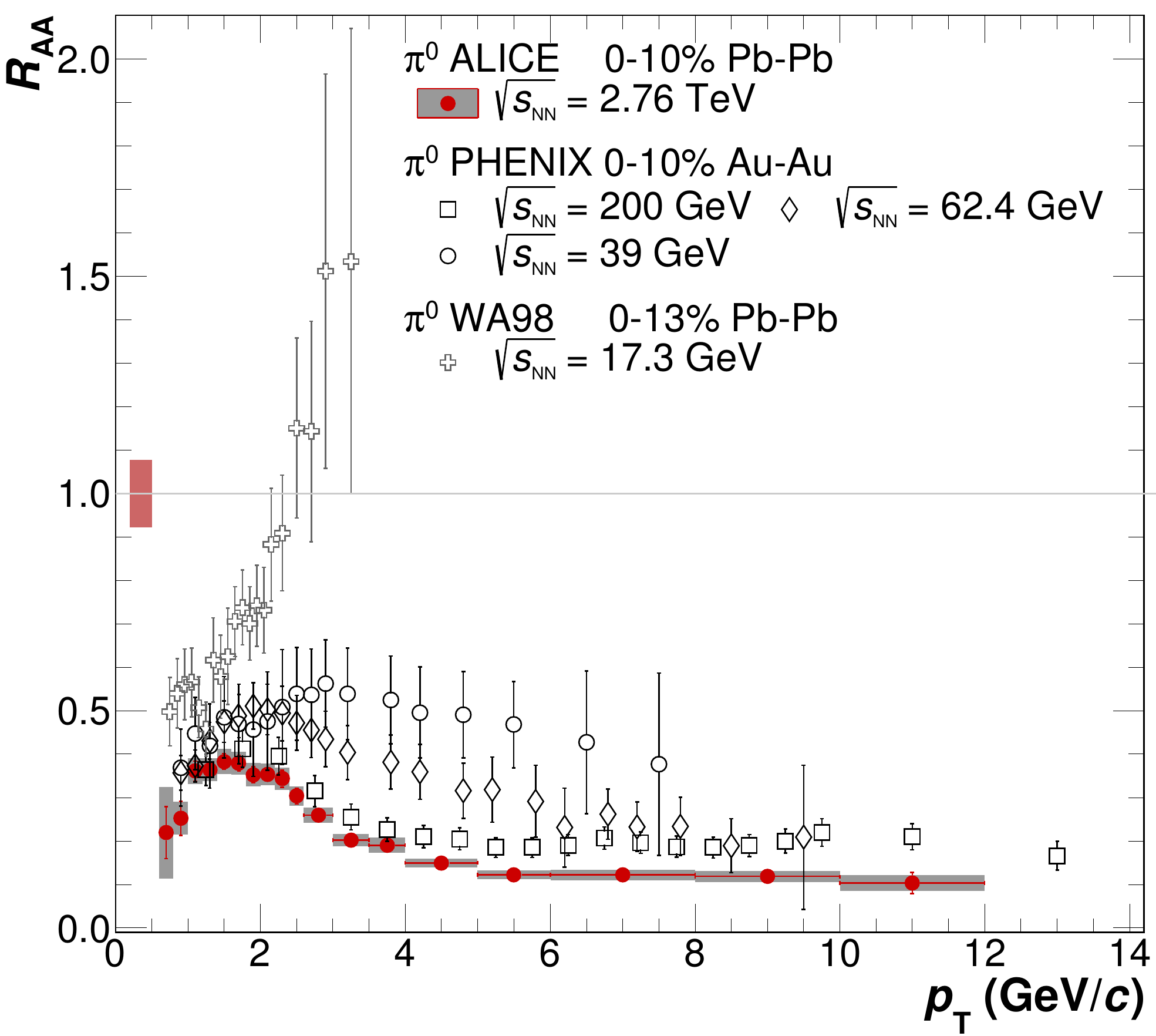}
\caption{(Color online) Neutral pion nuclear modification factor,
  \RAA, in \PbPb\
  collisions at $\snn = \unit[2.76]{TeV}$ for the $0-10\%$ class in
  comparison to results at lower energies.  The box around unity
  reflects the uncertainty of the average nuclear overlap function
  ($T_{AA}$) and the normalization uncertainty of the pp spectrum
  added in quadrature.  Horizontal bars reflect the bin width.  The
  center-of-mass energy dependence of the neutral pion \RAA\ is
  shown with results from Au--Au collisions at $\snn = 39$, $62.4$
  \cite{Adare:2012uk}, and $\unit[200]{GeV}$ \cite{Adare:2008qa} as
  well as the result from the CERN SPS \cite{Aggarwal:2007gw} (using
  scaled p-C data as reference) along with the results for \PbPb\ at
  $\snn = \unit[2.76]{TeV}$. The scale uncertainties of the
  measurements at lower energies of the order of $10-15\%$ are not
  shown.}
\label{fig:RAAother}

\end{figure}
\begin{figure}[htb]
  \centering
  \includegraphics[width=\figwidthnarrow]{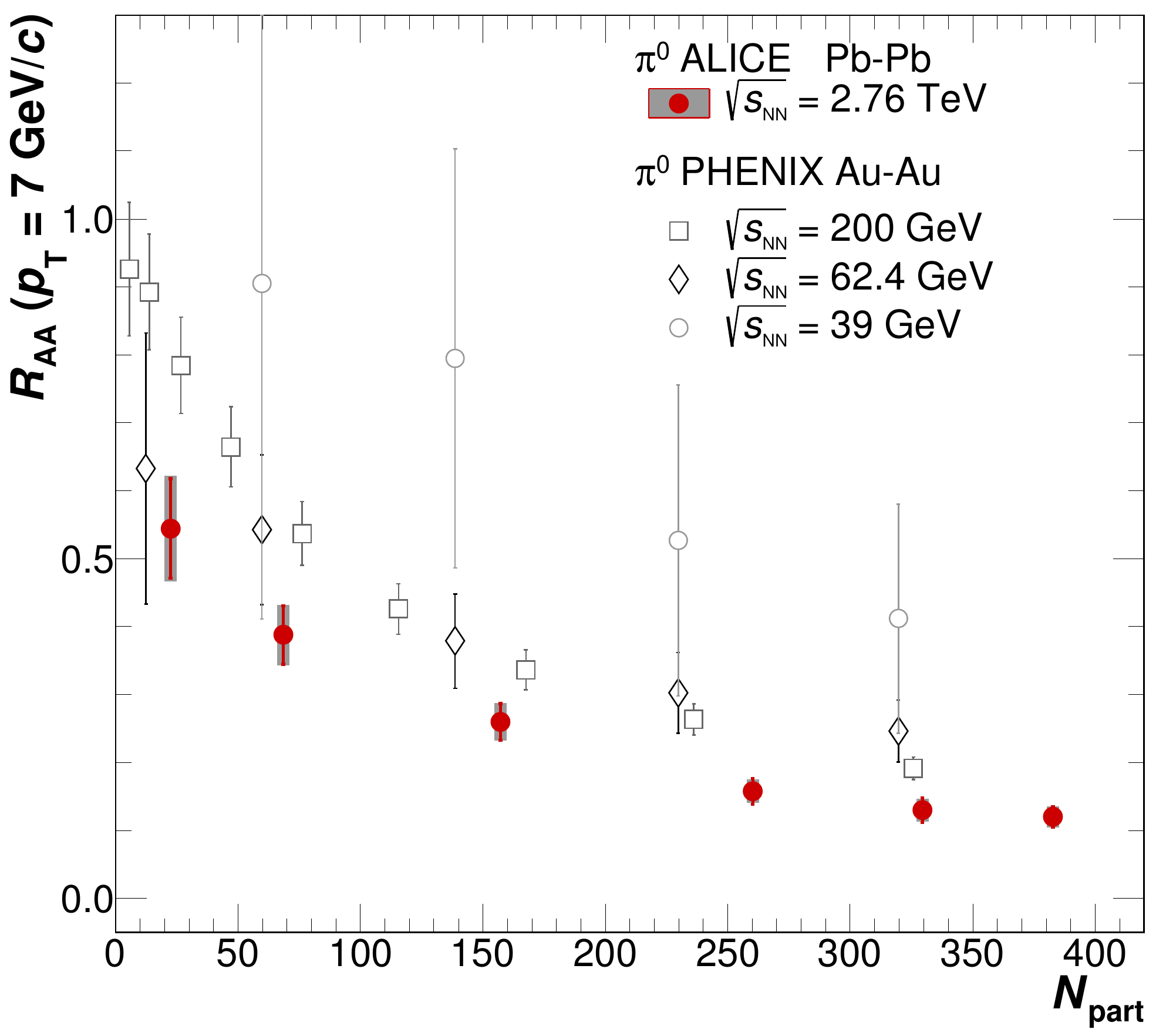}
  \caption{(Color online) Centrality dependence of the $\pi^0$ nuclear
    modification factor \RAA\ at $\pT = \unit[7]{GeV}/c$ in Au-Au and
    \PbPb\ collisions at $\snn = 39$, $62.4$, $200$
    \cite{Adare:2012uk,Adare:2012wg}, and
    $\unit[2760]{GeV}$.}
  \label{fig:RAA_vs_Npart}
\end{figure}

It is instructive to study the $\snn$ dependence of the neutral pion
\RAA. Fig.~\ref{fig:RAAother} shows that for central collisions the
\RAA\ at the LHC for $\pT \gtrsim \unit[2]{GeV}/c$ lies below the data
points at lower $\snn$. This indicates that the decrease of \RAA\
resulting from the higher initial energy densities created at larger
$\snn$ dominates over the increase of \RAA\ expected from the harder
initial parton $\pT$ spectra. To illustrated this point, one can
consider a somewhat oversimplified model with a $\pT$ independent
fractional energy loss $\varepsilon$ in conjunction with $\pT$ spectra
described by a power law \cite{Adler:2006bw}. In this model
$\varepsilon = 0.2$ corresponds to $\RAA^\mathrm{RHIC} \approx 0.25$
at $\snn = \unit[0.2]{TeV}$. The same fractional energy loss in
conjunction with the flatter spectra at $\snn = \unit[2.76]{TeV}$,
however, yield $\RAA^\mathrm{LHC} \approx 0.4$.  The shape of
$\RAA(\pT)$ in central collisions at $\snn = \unit[200]{GeV}$ and
$\snn = \unit[2.76]{TeV}$ appears to be similar. Considering the data
for all shown energies one observes that the value of \pT\ with the
maximum \RAA\ value appears to shift towards lower \pT\ with
increasing $\snn$. The centrality dependence of \RAA\ at $\pT =
\unit[7]{GeV}/c$ is shown in Fig.~\ref{fig:RAA_vs_Npart} for nuclear
collisions at $\snn = 39$, $62.4$, $200$
\cite{Adare:2012uk,Adare:2012wg}, and $\unit[2760]{GeV}$. At this
transverse momentum soft particle production from the bulk should be
negligible and parton energy loss is expected to be the dominant
effect. It can be seen that the suppression in \PbPb\ collisions at
the LHC is stronger than in Au-Au collisions at $\snn =
\unit[200]{GeV}$ for all centralities. In particular, the most
peripheral class of the LHC data already shows a sizable suppression
whereas at the lower energies the suppression appears to develop less
abruptly as a function of the number of participating nucleons
($\Npart$).

\begin{figure*}[htb]
\centering
\includegraphics[width=\figwidthwide]{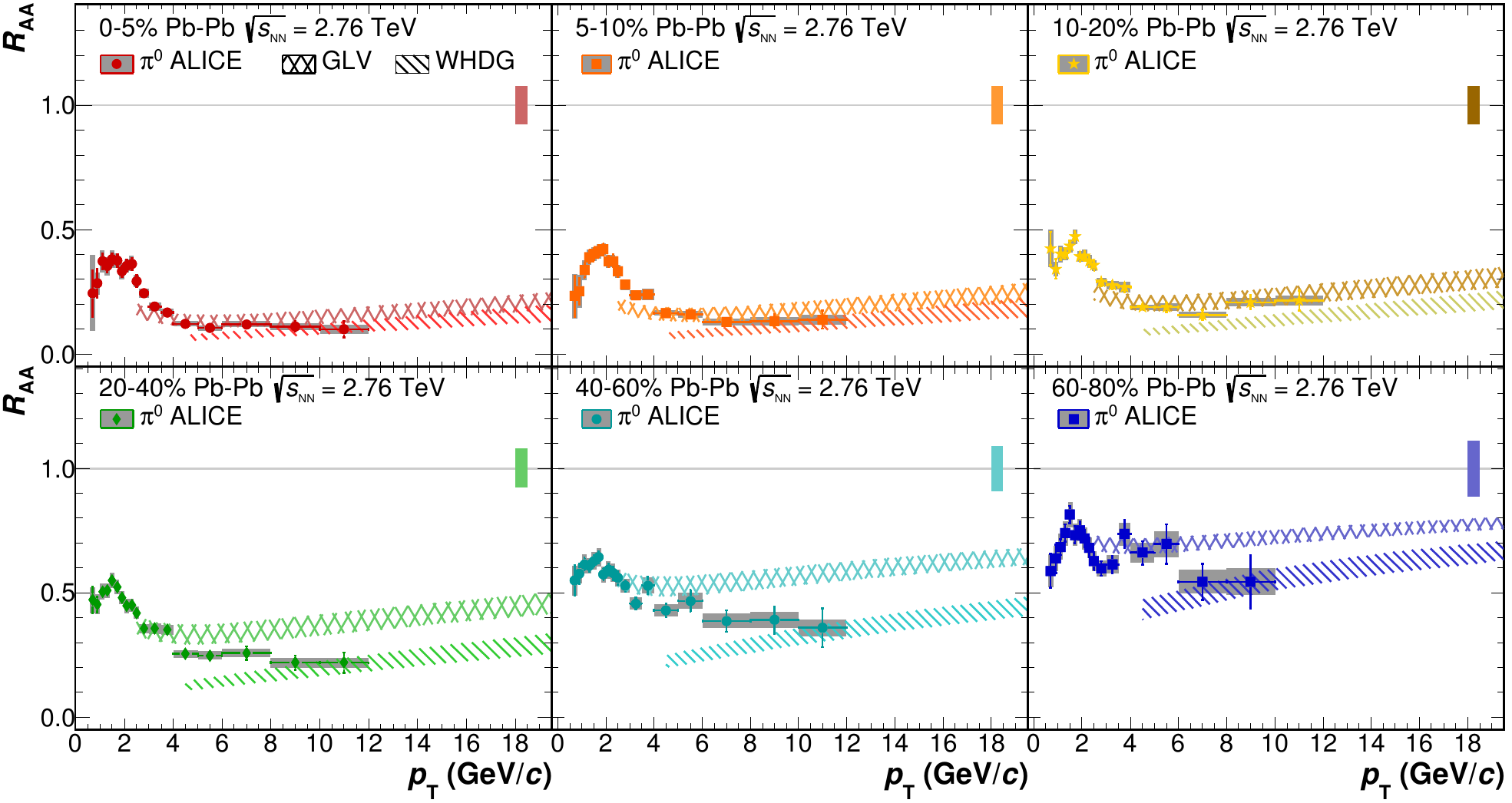}
\caption{(Color online) Comparison of the measured nuclear
  modification factor \RAA\ with a GLV calculation
  \cite{Sharma:2009hn,Neufeld:2010dz} and with a WHDG
  \cite{Horowitz:2007nq} parton energy loss calculations. Vertical
  lines show the statistical uncertainties, systematic uncertainties
  are shown as boxes. Horizontal lines indicate the bin width. The
  boxes around unity reflect the scale uncertainties of data related
  to $T_{AA}$ and the normalization of the pp spectrum.}
\label{fig:RAAtheory}
\end{figure*}

\begin{figure}[ht]
\centering
\includegraphics[width=\figwidthnarrow]{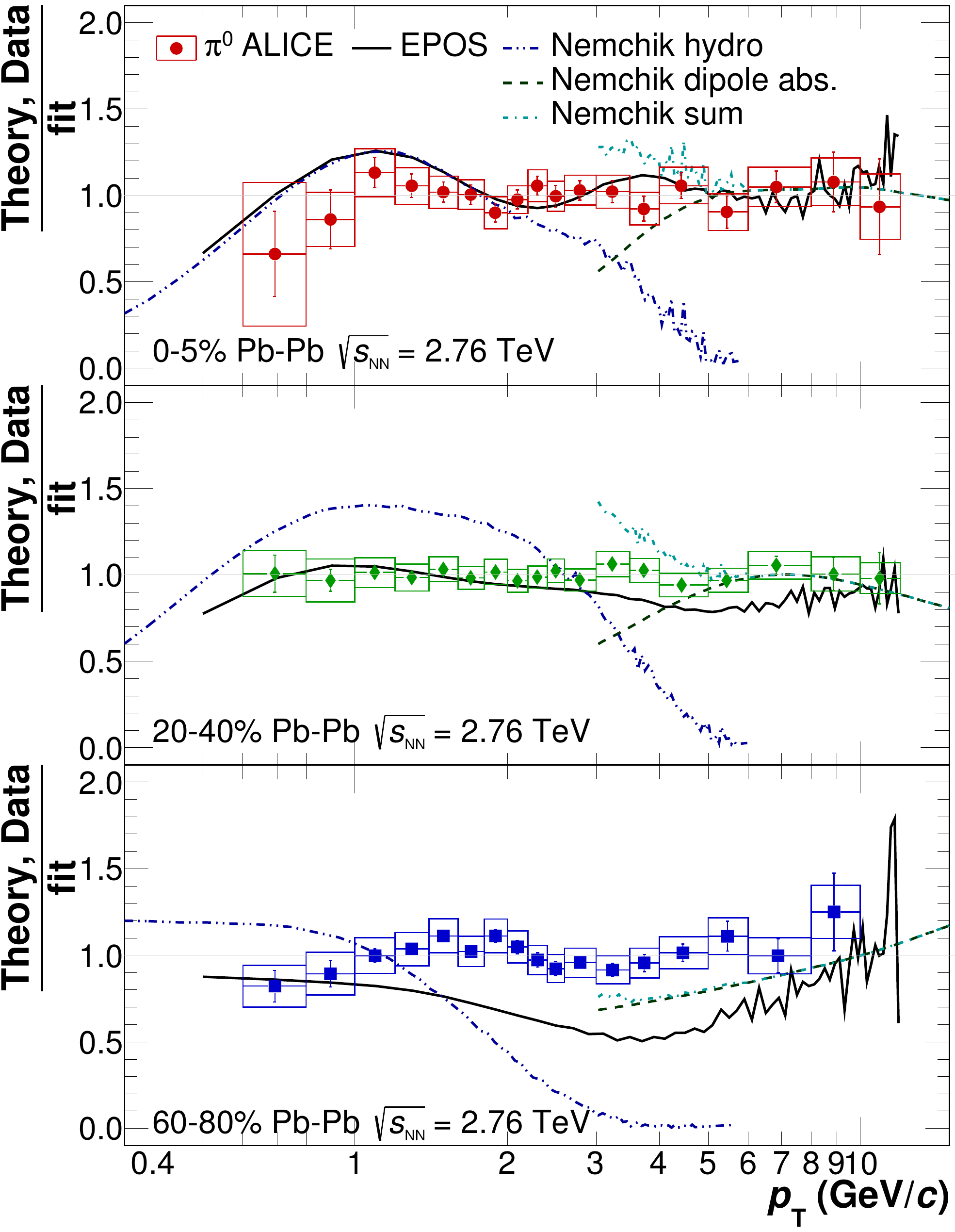}
\caption{(Color online) Comparison of the measured $\pi^0$ spectra for
  three centrality classes ($0-5\%$, $20-40\%$, $60-80\%$) with two
  calculations which make predictions for the full $\pT$ range of the
  measurement. The calculated spectra and the data points were divided
  by a fit of the measured $\pi^0$ spectra. For the data points the
  error bars represent the statistical uncertainties and the boxes the
  systematic uncertainties. Calculations with the EPOS event generator
  \cite{Werner:2012xh} are shown by the solid line. The fluctuations of the
  EPOS lines at high $\pT$ are due to limited statistics in the number of 
  generated events. The calculations
  by Nemchik et al.\ \cite{Kopeliovich:2012sc,Nemchik:2013ooa} combine
  a hydrodynamical model at low $\pT$ with a color dipole absorption
  model for $\pT \gtrsim \unit[3]{GeV}/c$. The two components and the
  sum (for $\pT \gtrsim \unit[3]{GeV}/c$) are shown separately.}
\label{fig:epos_and_dipole_absorp}
\end{figure}

In Fig.~\ref{fig:RAAtheory} the measured \RAA\ is compared with a GLV
model calculation \cite{Sharma:2009hn,Neufeld:2010dz} and with
theoretical predictions from the WHDG model \cite{Horowitz:2007nq}.
These models describe the interaction of a hard-scattered parton with
the medium of high color charge density within perturbative QCD
\cite{Armesto:2011ht}. Both calculations assume that the hadronization
of the hard-scattered parton occurs in the vacuum and is not affected
by the medium. They model the energy loss of the parton but not the
corresponding response of the medium. Their applicability is limited
to transverse momenta above $\unit[2-4]{GeV}/c$ as soft particle
production from the bulk is not taken into account. The \PbPb\ $\pi^0$
spectra are therefore also compared to two models which aim at a
description of the full $\pT$ range: an EPOS calculation
\cite{Werner:2012xh} and a calculation by Nemchik et al.\ based on the
combination of a hydrodynamic description at low $\pT$ and the
absorption of color dipoles at higher $\pT$
\cite{Kopeliovich:2012sc,Nemchik:2013ooa}. These comparisons are
presented in Fig.~\ref{fig:epos_and_dipole_absorp}.

The GLV calculation takes final-state radiative energy loss into
account. It includes the broadening of the transverse momenta of the
incoming partons in cold nuclear matter (``nuclear broadening'' or
``Cronin effect''). The main parameter of this model, the initial
gluon density, was tuned to describe the neutral pion suppression
observed in Au-Au collisions at RHIC. For the calculation of the
parton energy loss in \PbPb\ collisions at the LHC the initial gluon
density was constrained by the measured charged-particle
multi\-pli\-ci\-ties.  The model can approximately reproduce the
centrality and $\pT$ dependence of the $\pi^0$ \RAA.

The WHDG model takes into account collisional and radiative parton
energy loss and geometrical path length fluctuations. The color charge
density of the medi\-um is assumed to be proportional to the number of
participating nucleons from a Glauber model, and hard parton-parton
scatterings are proportional to the number of binary nucleon-nucleon
collisions. Parameters of the model were constrained by the neutral
pion \RAA\ measured at RHIC. Like in the case of the GLV calculation,
the neutral pion \RAA\ at the LHC is then predicted by translating the
measured charged-particle multiplicity
$\mathrm{d}N_{ch}/\mathrm{d}\eta$ in \PbPb\ collisions into an initial
gluon density which is the free parameter of the model. For central
collisions this yielded an increase in the gluon density from
$\mathrm{d}N_g/\mathrm{d}y \approx 1400$ at RHIC to
$\mathrm{d}N_g/\mathrm{d}y \approx 3000$ at the LHC. The WHDG model
reproduces the $\pi^0$ \RAA\ in central collisions reasonably well,
but predicts too strong suppression for more peripheral classes.

The two model predictions for the full $\pT$ range are compared to the
measured spectra in Fig.~\ref{fig:epos_and_dipole_absorp}. EPOS is
based on the hadronization of flux tubes produced early in the
collision. Hard scattering in this model produces strings with
transversely moving parts. String segments with low energies are
assumed to be part of the bulk whose space-time evolution is modeled
within hydrodynamics.  String segments with sufficiently large energy
fragment in the vacuum. A third class of string segments with
intermediate energies is considered to have enough energy to leave the
medium accompanied by quark pick-up from the bulk during the
fragmentation process. In EPOS particle production is determined by
hydrodynamic flow at low $\pT\ (\lesssim \unit[4]{GeV}/c)$, followed
at higher $\pT$ by energy loss of high-$\pT$ string segments. In
central collisions the EPOS calculation describes the measured $\pi^0$
spectrum rather well. Towards more peripheral collisions a discrepancy
develops for $1 \lesssim \pT \lesssim \unit[5]{GeV}/c$ which may
possibly be attributed to underestimating the contribution of
hydrodynamic flow in peripheral collisions.

The calculation by Nemchik et al.\ also combines a model for hadron
suppression at high $\pT$ with a hydrodynamic description of bulk
particle production at low $\pT$.  Hadron suppression in this model
results from the absorption of pre-hadrons, i.e., of color dipoles
which are already formed in the medium by hard-scattered partons
during the production of hadrons with large $z =
p_\mathrm{hadron}/p_\mathrm{parton}$. As the model, at high $\pT$,
predicts only \RAA, the calculated \RAA\ values were scaled by
$\langle T_\mathrm{AA} \rangle \times
E\,\mathrm{d}^3\sigma_\mathrm{meas}^{\pi^0}/\mathrm{d}^3p$ and then
added to the calculated $\pi^0$ invariant yields from the hydrodynamic
model in order to compare to the measured $\pi^0$ spectra. The
hydrodynamic calculation dominates the total $\pi^0$ yield up to $\pT
= \unit[2]{GeV}/c$ and remains a significant contribution up to
$\unit[5]{GeV}/c$. From about $\unit[3]{GeV}/c$ the contribution from
hard scattering becomes larger than the one from the hydrodynamic
calculation. The spectrum in central \PbPb\ collisions ($0-5\%$) is
approximately described except for the transition region between the
hydrodynamic and the hard contribution.  In the $20-40\%$ class the
hydrodynamic calculation overpredicts the data up to $\pT =
\unit[2]{GeV}/c$.

%%% Local Variables: 
%%% mode: latex
%%% TeX-master: "ALICE_PbPb2pi0"
%%% End: 

\section{Conclusions}
\label{sec:Conclusions}

Measurements of neutral pion production at midrapidity in \pp\ and
\PbPb\ collisions at $\snn=\unit[2.76]{TeV}$ were presented. The
measurements were performed with two independent techniques, by
measuring the photons with the electromagnetic calorimeter PHOS, and
by measuring converted photons with the ALICE tracking system. The two
independent measurements were found to give consistent results, and
were combined for the final results.

The neutral pion spectrum in pp collisions was compared to a NLO
perturbative QCD calculation using the DSS fragmentation
functions. This calculation, which describes the pion spectrum in pp
collisions at $\sqrt{s} = \unit[0.9]{TeV}$ rather well, tends to
overpredict the $\pi^0$ cross section already at $\sqrt{s} =
\unit[2.76]{TeV}$. Along with a similar observation in pp collision at
$\sqrt{s} = \unit[7]{TeV}$ this indicates the likely need for
improvements in the gluon-to-pion fragmentation function. A similar
observation was made for transverse momentum spectra of charged
particles in proton-proton and proton-antiproton collisons at $1.96
\lesssim \sqrt{s} \lesssim \unit[7]{TeV}$
\cite{d'Enterria:2013vba,Abelev:2013ala}.

The neutral pion nuclear suppression factor \RAA\ was calculated from
the measured neutral pion spectra, and was compared to measurements at
lower energies and to theoretical predictions. The $\pi^0$ suppression
in the most central class ($0-5\%$) reaches values of up to $8-10$ for
$5 \lesssim \pT \lesssim \unit[7]{GeV}/c$. The suppression in \PbPb\
collisions at $\snn = \unit[2.76]{TeV}$ is stronger than in Au-Au
collisions at $\snn = \unit[200]{GeV}$ (and lower energies) at RHIC for
all centralities.

The general features of the centrality and $\pT$ dependence of the
\RAA\ for $\pT \gtrsim \unit[2]{GeV}/c$ are approximately reproduced
by GLV and WHDG parton energy loss calculations, although the WHDG
calculation performs less well in peripheral collisions. For both
calculations the main free parameter, the initial gluon density, was
chosen to describe the neutral pion suppression at RHIC and then
scaled to LHC energies based on the measured charged-particle
multiplicities. The measured $\pi^0$ spectra were also compared to
calculations with the EPOS event generator and a calculation by
Nemchik et al.  By combining soft particle production from a
hydrodynamically evolving medium with a model for hadron suppression
these models are capable of making predictions for the entire $\pT$
range. An important task on the theoretical side will be to establish
whether the observed deviations from the data simply indicate a
suboptimal adjustment of parameters or hint at important physical
phenomena missing in the models.  Future analyses based on runs with
higher integrated luminosities, e.g. the 2011 LHC Pb-Pb run, will also
include the ALICE lead-scintillator electromagnetic calorimeter
(EMCal) and will allow us to extend the neutral pion measurement to
higher transverse momenta. The role of initial-state effects on the
particle production in Pb-Pb collisions will be investigated by
measurements of particle production in p-Pb collisions.

%%% Local Variables: 
%%% mode: latex
%%% TeX-master: "ALICE_PbPb2pi0"
%%% End: 

\section*{Acknowledgements}
We would like to thank  Jan Nemchik, William A.\ Horowitz, Ivan Vitev, and Klaus Werner 
for providing the model calculations shown in this paper. This work was supported 
by the grants RFBR~10-02-91052 and RFBR 12-02-91527.
The ALICE Collaboration would like to thank all its engineers and technicians for their invaluable contributions to the construction of the experiment and the CERN accelerator teams for the outstanding performance of the LHC complex.
%\\
The ALICE Collaboration gratefully acknowledges the resources and support provided by all Grid centres and the Worldwide LHC Computing Grid (WLCG) collaboration.
%\\
The ALICE Collaboration acknowledges the following funding agencies for their support in building and
running the ALICE detector:
 %\\
State Committee of Science,  World Federation of Scientists (WFS)
and Swiss Fonds Kidagan, Armenia,
 %\\
Conselho Nacional de Desenvolvimento Cient\'{\i}fico e Tecnol\'{o}gico (CNPq), Financiadora de Estudos e Projetos (FINEP),
Funda\c{c}\~{a}o de Amparo \`{a} Pesquisa do Estado de S\~{a}o Paulo (FAPESP);
 %\\
National Natural Science Foundation of China (NSFC), the Chinese Ministry of Education (CMOE)
and the Ministry of Science and Technology of China (MSTC);
 %\\
Ministry of Education and Youth of the Czech Republic;
 %\\
Danish Natural Science Research Council, the Carlsberg Foundation and the Danish National Research Foundation;
 %\\
The European Research Council under the European Community's Seventh Framework Programme;
 %\\
Helsinki Institute of Physics and the Academy of Finland;
 %\\
French CNRS-IN2P3, the `Region Pays de Loire', `Region Alsace', `Region Auvergne' and CEA, France;
 %\\
German BMBF and the Helmholtz Association;
%\\
General Secretariat for Research and Technology, Ministry of
Development, Greece;
%\\
Hungarian OTKA and National Office for Research and Technology (NKTH);
 %\\
Department of Atomic Energy and Department of Science and Technology of the Government of India;
 %\\
Istituto Nazionale di Fisica Nucleare (INFN) and Centro Fermi -
Museo Storico della Fisica e Centro Studi e Ricerche "Enrico
Fermi", Italy;
 %\\
MEXT Grant-in-Aid for Specially Promoted Research, Ja\-pan;
 %\\
Joint Institute for Nuclear Research, Dubna;
 %\\
%Korea Foundation for International Cooperation of Science and Technology (KICOS);
National Research Foundation of Korea (NRF);
 %\\
CONACYT, DGAPA, M\'{e}xico, ALFA-EC and the EPLANET Program
(European Particle Physics Latin American Network)
 %\\
Stichting voor Fundamenteel Onderzoek der Materie (FOM) and the Nederlandse Organisatie voor Wetenschappelijk Onderzoek (NWO), Netherlands;
 %\\
Research Council of Norway (NFR);
 %\\
Polish Ministry of Science and Higher Education;
 %\\
National Science Centre, Poland;
 %\\
 Ministry of National Education/Institute for Atomic Physics and CNCS-UEFISCDI - Romania;
 %\\
Ministry of Education and Science of Russian Federation, Russian
Academy of Sciences, Russian Federal Agency of Atomic Energy,
Russian Federal Agency for Science and Innovations and The Russian
Foundation for Basic Research;
 %\\
Ministry of Education of Slovakia;
 %\\
Department of Science and Technology, South Africa;
 %\\
CIEMAT, EELA, Ministerio de Econom\'{i}a y Competitividad (MINECO) of Spain, Xunta de Galicia (Conseller\'{\i}a de Educaci\'{o}n),
CEA\-DEN, Cubaenerg\'{\i}a, Cuba, and IAEA (International Atomic Energy Agency);
 %\\
Swedish Research Council (VR) and Knut $\&$ Alice Wallenberg
Foundation (KAW);
 %\\
Ukraine Ministry of Education and Science;
 %\\
United Kingdom Science and Technology Facilities Council (STFC);
 %\\
The United States Department of Energy, the United States National
Science Foundation, the State of Texas, and the State of Ohio.

%\clearpage
%\bibliographystyle{bibstyle}
% \bibliographystyle{apsrev4-1} % Phys Rev bibtex style
\bibliographystyle{utphys} % Phys Rev bibtex style
\bibliography{pi0raa.bib}

\providecommand{\href}[2]{#2}\begingroup\raggedright\begin{thebibliography}{10}

\bibitem{Borsanyi:2010cj}
S.~Borsanyi, G.~Endrodi, Z.~Fodor, A.~Jakovac, S.~D. Katz, {\em et~al.}, ``{The
  QCD equation of state with dynamical quarks},''
  \href{http://dx.doi.org/10.1007/JHEP11(2010)077}{{\em JHEP} {\bfseries 1011}
  (2010) 077},
\href{http://arxiv.org/abs/1007.2580}{{\ttfamily arXiv:1007.2580 [hep-lat]}}.
%%CITATION = ARXIV:1007.2580;%%.

\bibitem{Bazavov:2011nk}
A.~Bazavov, T.~Bhattacharya, M.~Cheng, C.~DeTar, H.~Ding, {\em et~al.}, ``{The
  chiral and deconfinement aspects of the QCD transition},''
  \href{http://dx.doi.org/10.1103/PhysRevD.85.054503}{{\em Phys.Rev.}
  {\bfseries D85} (2012) 054503},
\href{http://arxiv.org/abs/1111.1710}{{\ttfamily arXiv:1111.1710 [hep-lat]}}.
%%CITATION = ARXIV:1111.1710;%%.

\bibitem{Chatrchyan:2012mb}
{\bfseries CMS} Collaboration, S.~Chatrchyan {\em et~al.}, ``{Measurement of
  the pseudorapidity and centrality dependence of the transverse energy density
  in PbPb collisions at $\sqrt{s_{NN}}=2.76$ TeV},''
  \href{http://dx.doi.org/10.1103/PhysRevLett.109.152303}{{\em Phys.Rev.Lett.}
  {\bfseries 109} (2012) 152303},
\href{http://arxiv.org/abs/1205.2488}{{\ttfamily arXiv:1205.2488 [nucl-ex]}}.
%%CITATION = ARXIV:1205.2488;%%.

\bibitem{Alice:2011rta}
A.~Toia, ``{Bulk Properties of Pb-Pb collisions at sqrt(sNN) = 2.76 TeV
  measured by ALICE},''
  \href{http://dx.doi.org/10.1088/0954-3899/38/12/124007}{{\em J.Phys.G}
  {\bfseries G38} (2011) 124007},
\href{http://arxiv.org/abs/1107.1973}{{\ttfamily arXiv:1107.1973 [nucl-ex]}}.
%%CITATION = ARXIV:1107.1973;%%.

\bibitem{Heinz:2013th}
U.~Heinz and R.~Snellings, ``{Collective flow and viscosity in relativistic
  heavy-ion collisions},''
  \href{http://dx.doi.org/10.1146/annurev-nucl-102212-170540}{{\em
  Ann.Rev.Nucl.Part.Sci.} {\bfseries 63} (2013) 123--151},
\href{http://arxiv.org/abs/1301.2826}{{\ttfamily arXiv:1301.2826 [nucl-th]}}.
%%CITATION = ARXIV:1301.2826;%%.

\bibitem{Bjorken:1982tu}
J.~Bjorken, ``{Energy Loss of Energetic Partons in Quark - Gluon Plasma:
  Possible Extinction of High p(t) Jets in Hadron - Hadron Collisions},''
\href{http://arxiv.org/abs/FERMILAB-PUB-82-059-THY,
  FERMILAB-PUB-82-059-T}{{\ttfamily FERMILAB-PUB-82-059-THY,
  FERMILAB-PUB-82-059-T}}.
%%CITATION = FERMILAB-PUB-82-059-THY ETC.;%%.

\bibitem{Wang:1991xy}
X.-N. Wang and M.~Gyulassy, ``{Gluon shadowing and jet quenching in A + A
  collisions at s**(1/2) = 200-GeV},''
\href{http://dx.doi.org/10.1103/PhysRevLett.68.1480}{{\em Phys.Rev.Lett.}
  {\bfseries 68} (1992) 1480--1483}.
%%CITATION = PRLTA,68,1480;%%.

\bibitem{Wiedemann:2009sh}
U.~A. Wiedemann, \href{http://dx.doi.org/10.1007/978-3-642-01539-7\_17}{``{Jet
  Quenching in Heavy Ion Collisions},''} in {\em SpringerMaterials - The
  Landolt-B{\"o}rnstein Database}, R.~Stock, ed., vol.~23: Relativistic Heavy
  Ion Physics.
\newblock Springer-Verlag Berlin Heidelberg, 2009.
\newblock
\href{http://arxiv.org/abs/0908.2306}{{\ttfamily arXiv:0908.2306 [hep-ph]}}.
\newblock
%%CITATION = ARXIV:0908.2306;%%.

\bibitem{d'Enterria:2009am}
D.~d'Enterria, \href{http://dx.doi.org/10.1007/978-3-642-01539-7\_16}{``Jet
  quenching,''} in {\em SpringerMaterials - The Landolt-B{\"o}rnstein
  Database}, R.~Stock, ed., vol.~23: Relativistic Heavy Ion Physics.
\newblock Springer-Verlag Berlin Heidelberg, 2009.
\newblock
\href{http://arxiv.org/abs/0902.2011}{{\ttfamily arXiv:0902.2011 [nucl-ex]}}.
\newblock
%%CITATION = ARXIV:0902.2011;%%.

\bibitem{Majumder:2010qh}
A.~Majumder and M.~Van~Leeuwen, ``{The Theory and Phenomenology of Perturbative
  QCD Based Jet Quenching},''
  \href{http://dx.doi.org/10.1016/j.ppnp.2010.09.001}{{\em
  Prog.Part.Nucl.Phys.} {\bfseries A66} (2011) 41--92},
\href{http://arxiv.org/abs/1002.2206}{{\ttfamily arXiv:1002.2206 [hep-ph]}}.
%%CITATION = ARXIV:1002.2206;%%.

\bibitem{Armesto:2011ht}
N.~Armesto, B.~Cole, C.~Gale, W.~A. Horowitz, P.~Jacobs, {\em et~al.},
  ``{Comparison of Jet Quenching Formalisms for a Quark-Gluon Plasma
  'Brick'},'' \href{http://dx.doi.org/10.1103/PhysRevC.86.064904}{{\em
  Phys.Rev.} {\bfseries C86} (2012) 064904},
\href{http://arxiv.org/abs/1106.1106}{{\ttfamily arXiv:1106.1106 [hep-ph]}}.
%%CITATION = ARXIV:1106.1106;%%.

\bibitem{Burke:2013yra}
K.~M. Burke, A.~Buzzatti, N.~Chang, C.~Gale, M.~Gyulassy, {\em et~al.},
  ``{Extracting jet transport coefficient from jet quenching at RHIC and
  LHC},''
\href{http://arxiv.org/abs/1312.5003}{{\ttfamily arXiv:1312.5003 [nucl-th]}}.
%%CITATION = ARXIV:1312.5003;%%.

\bibitem{ALICE:2012mj}
{\bfseries ALICE} Collaboration, B.~Abelev {\em et~al.}, ``{Transverse Momentum
  Distribution and Nuclear Modification Factor of Charged Particles in $p$-Pb
  Collisions at $\sqrt{s_{NN}}=5.02$ TeV},''
  \href{http://dx.doi.org/10.1103/PhysRevLett.110.082302}{{\em Phys.Rev.Lett.}
  {\bfseries 110} (2013) 082302},
\href{http://arxiv.org/abs/1210.4520}{{\ttfamily arXiv:1210.4520 [nucl-ex]}}.
%%CITATION = ARXIV:1210.4520;%%.

\bibitem{Horowitz:2011gd}
W.~Horowitz and M.~Gyulassy, ``{The Surprising Transparency of the sQGP at
  LHC},'' \href{http://dx.doi.org/10.1016/j.nuclphysa.2011.09.018}{{\em
  Nucl.Phys.} {\bfseries A872} (2011) 265--285},
\href{http://arxiv.org/abs/1104.4958}{{\ttfamily arXiv:1104.4958 [hep-ph]}}.
%%CITATION = ARXIV:1104.4958;%%.

\bibitem{Sassot:2010bh}
R.~Sassot, P.~Zurita, and M.~Stratmann, ``{Inclusive Hadron Production in the
  CERN-LHC Era},'' \href{http://dx.doi.org/10.1103/PhysRevD.82.074011}{{\em
  Phys.Rev.} {\bfseries D82} (2010) 074011},
\href{http://arxiv.org/abs/1008.0540}{{\ttfamily arXiv:1008.0540 [hep-ph]}}.
%%CITATION = ARXIV:1008.0540;%%.

\bibitem{Sassot:2009sh}
R.~Sassot, M.~Stratmann, and P.~Zurita, ``{Fragmentations Functions in Nuclear
  Media},'' \href{http://dx.doi.org/10.1103/PhysRevD.81.054001}{{\em Phys.Rev.}
  {\bfseries D81} (2010) 054001},
\href{http://arxiv.org/abs/0912.1311}{{\ttfamily arXiv:0912.1311 [hep-ph]}}.
%%CITATION = ARXIV:0912.1311;%%.

\bibitem{Sapeta:2007ad}
S.~Sapeta and U.~A. Wiedemann, ``{Jet hadrochemistry as a characteristics of
  jet quenching},''
  \href{http://dx.doi.org/10.1140/epjc/s10052-008-0592-8}{{\em Eur.Phys.J.}
  {\bfseries C55} (2008) 293--302},
\href{http://arxiv.org/abs/0707.3494}{{\ttfamily arXiv:0707.3494 [hep-ph]}}.
%%CITATION = ARXIV:0707.3494;%%.

\bibitem{Bellwied:2010pr}
R.~Bellwied and C.~Markert, ``{In-medium hadronization in the deconfined matter
  at RHIC and LHC},''
  \href{http://dx.doi.org/10.1016/j.physletb.2010.06.028}{{\em Phys.Lett.}
  {\bfseries B691} (2010) 208--213},
\href{http://arxiv.org/abs/1005.5416}{{\ttfamily arXiv:1005.5416 [nucl-th]}}.
%%CITATION = ARXIV:1005.5416;%%.

\bibitem{Adcox:2001jp}
{\bfseries PHENIX} Collaboration, K.~Adcox {\em et~al.}, ``{Suppression of
  hadrons with large transverse momentum in central Au+Au collisions at
  $\sqrt{s_{NN}}$ = 130-GeV},''
  \href{http://dx.doi.org/10.1103/PhysRevLett.88.022301}{{\em Phys.Rev.Lett.}
  {\bfseries 88} (2002) 022301},
\href{http://arxiv.org/abs/nucl-ex/0109003}{{\ttfamily arXiv:nucl-ex/0109003
  [nucl-ex]}}.
%%CITATION = NUCL-EX/0109003;%%.

\bibitem{Adler:2002xw}
{\bfseries STAR} Collaboration, C.~Adler {\em et~al.}, ``{Centrality dependence
  of high $p_{T}$ hadron suppression in Au+Au collisions at $\sqrt{s}_{NN}$ =
  130-GeV},'' \href{http://dx.doi.org/10.1103/PhysRevLett.89.202301}{{\em
  Phys.Rev.Lett.} {\bfseries 89} (2002) 202301},
\href{http://arxiv.org/abs/nucl-ex/0206011}{{\ttfamily arXiv:nucl-ex/0206011
  [nucl-ex]}}.
%%CITATION = NUCL-EX/0206011;%%.

\bibitem{Agakishiev:2011dc}
{\bfseries STAR} Collaboration, G.~Agakishiev {\em et~al.}, ``{Identified
  hadron compositions in p+p and Au+Au collisions at high transverse momenta at
  $\sqrt{s_{_{NN}}} = 200$ GeV},''
  \href{http://dx.doi.org/10.1103/PhysRevLett.108.072302}{{\em Phys.Rev.Lett.}
  {\bfseries 108} (2012) 072302},
\href{http://arxiv.org/abs/1110.0579}{{\ttfamily arXiv:1110.0579 [nucl-ex]}}.
%%CITATION = ARXIV:1110.0579;%%.

\bibitem{Adare:2012wg}
{\bfseries PHENIX} Collaboration, A.~Adare {\em et~al.}, ``{Neutral pion
  production with respect to centrality and reaction plane in Au$+$Au
  collisions at $\sqrt{s_{NN}}$=200 GeV},''
  \href{http://dx.doi.org/10.1103/PhysRevC.87.034911}{{\em Phys.Rev.}
  {\bfseries C87} (2013) 034911},
\href{http://arxiv.org/abs/1208.2254}{{\ttfamily arXiv:1208.2254 [nucl-ex]}}.
%%CITATION = ARXIV:1208.2254;%%.

\bibitem{Adare:2013esx}
{\bfseries PHENIX} Collaboration, A.~Adare {\em et~al.}, ``{Spectra and ratios
  of identified particles in Au+Au and d+Au collisions at $\sqrt{s_{NN}}=200$
  GeV},'' \href{http://dx.doi.org/10.1103/PhysRevC.88.024906}{{\em Phys.Rev.}
  {\bfseries C88} (2013) 024906},
\href{http://arxiv.org/abs/1304.3410}{{\ttfamily arXiv:1304.3410 [nucl-ex]}}.
%%CITATION = ARXIV:1304.3410;%%.

\bibitem{Adler:2002tq}
{\bfseries STAR} Collaboration, C.~Adler {\em et~al.}, ``{Disappearance of
  back-to-back high $p_{T}$ hadron correlations in central Au+Au collisions at
  $\sqrt{s_{NN}}$ = 200-GeV},''
  \href{http://dx.doi.org/10.1103/PhysRevLett.90.082302}{{\em Phys.Rev.Lett.}
  {\bfseries 90} (2003) 082302},
\href{http://arxiv.org/abs/nucl-ex/0210033}{{\ttfamily arXiv:nucl-ex/0210033
  [nucl-ex]}}.
%%CITATION = NUCL-EX/0210033;%%.

\bibitem{Adams:2006yt}
{\bfseries STAR} Collaboration, J.~Adams {\em et~al.}, ``{Direct observation of
  dijets in central Au+Au collisions at s(NN)**(1/2) = 200-GeV},''
  \href{http://dx.doi.org/10.1103/PhysRevLett.97.162301}{{\em Phys.Rev.Lett.}
  {\bfseries 97} (2006) 162301},
\href{http://arxiv.org/abs/nucl-ex/0604018}{{\ttfamily arXiv:nucl-ex/0604018
  [nucl-ex]}}.
%%CITATION = NUCL-EX/0604018;%%.

\bibitem{Arsene:2004fa}
{\bfseries BRAHMS} Collaboration, I.~Arsene {\em et~al.}, ``{Quark gluon plasma
  and color glass condensate at RHIC? The Perspective from the BRAHMS
  experiment},'' \href{http://dx.doi.org/10.1016/j.nuclphysa.2005.02.130}{{\em
  Nucl.Phys.} {\bfseries A757} (2005) 1--27},
\href{http://arxiv.org/abs/nucl-ex/0410020}{{\ttfamily arXiv:nucl-ex/0410020
  [nucl-ex]}}.
%%CITATION = NUCL-EX/0410020;%%.

\bibitem{Adcox:2004mh}
{\bfseries PHENIX} Collaboration, K.~Adcox {\em et~al.}, ``{Formation of dense
  partonic matter in relativistic nucleus-nucleus collisions at RHIC:
  Experimental evaluation by the PHENIX collaboration},''
  \href{http://dx.doi.org/10.1016/j.nuclphysa.2005.03.086}{{\em Nucl.Phys.}
  {\bfseries A757} (2005) 184--283},
\href{http://arxiv.org/abs/nucl-ex/0410003}{{\ttfamily arXiv:nucl-ex/0410003
  [nucl-ex]}}.
%%CITATION = NUCL-EX/0410003;%%.

\bibitem{Back:2004je}
B.~Back, M.~Baker, M.~Ballintijn, D.~Barton, B.~Becker, {\em et~al.}, ``{The
  PHOBOS perspective on discoveries at RHIC},''
  \href{http://dx.doi.org/10.1016/j.nuclphysa.2005.03.084}{{\em Nucl.Phys.}
  {\bfseries A757} (2005) 28--101},
\href{http://arxiv.org/abs/nucl-ex/0410022}{{\ttfamily arXiv:nucl-ex/0410022
  [nucl-ex]}}.
%%CITATION = NUCL-EX/0410022;%%.

\bibitem{Adams:2005dq}
{\bfseries STAR} Collaboration, J.~Adams {\em et~al.}, ``{Experimental and
  theoretical challenges in the search for the quark gluon plasma: The STAR
  Collaboration's critical assessment of the evidence from RHIC collisions},''
  \href{http://dx.doi.org/10.1016/j.nuclphysa.2005.03.085}{{\em Nucl.Phys.}
  {\bfseries A757} (2005) 102--183},
\href{http://arxiv.org/abs/nucl-ex/0501009}{{\ttfamily arXiv:nucl-ex/0501009
  [nucl-ex]}}.
%%CITATION = NUCL-EX/0501009;%%.

\bibitem{Adler:2003qi}
{\bfseries PHENIX} Collaboration, S.~Adler {\em et~al.}, ``{Suppressed pi0
  production at large transverse momentum in central Au+ Au collisions at
  S(NN)**1/2 = 200 GeV},''
  \href{http://dx.doi.org/10.1103/PhysRevLett.91.072301}{{\em Phys.Rev.Lett.}
  {\bfseries 91} (2003) 072301},
\href{http://arxiv.org/abs/nucl-ex/0304022}{{\ttfamily arXiv:nucl-ex/0304022
  [nucl-ex]}}.
%%CITATION = NUCL-EX/0304022;%%.

\bibitem{Adare:2008qa}
{\bfseries PHENIX} Collaboration, A.~Adare {\em et~al.}, ``{Suppression pattern
  of neutral pions at high transverse momentum in Au + Au collisions at
  s(NN)**(1/2) = 200-GeV and constraints on medium transport coefficients},''
  \href{http://dx.doi.org/10.1103/PhysRevLett.101.232301}{{\em Phys.Rev.Lett.}
  {\bfseries 101} (2008) 232301},
\href{http://arxiv.org/abs/0801.4020}{{\ttfamily arXiv:0801.4020 [nucl-ex]}}.
%%CITATION = ARXIV:0801.4020;%%.

\bibitem{Bass:2008rv}
S.~A. Bass, C.~Gale, A.~Majumder, C.~Nonaka, G.-Y. Qin, {\em et~al.},
  ``{Systematic Comparison of Jet Energy-Loss Schemes in a realistic
  hydrodynamic medium},''
  \href{http://dx.doi.org/10.1103/PhysRevC.79.024901}{{\em Phys.Rev.}
  {\bfseries C79} (2009) 024901},
\href{http://arxiv.org/abs/0808.0908}{{\ttfamily arXiv:0808.0908 [nucl-th]}}.
%%CITATION = ARXIV:0808.0908;%%.

\bibitem{Adare:2008ad}
{\bfseries PHENIX} Collaboration, A.~Adare {\em et~al.}, ``{Onset of pi0
  Suppression Studied in Cu+Cu Collisions at sNN=22.4, 62.4, and 200 GeV},''
  \href{http://dx.doi.org/10.1103/PhysRevLett.101.162301}{{\em Phys.Rev.Lett.}
  {\bfseries 101} (2008) 162301},
\href{http://arxiv.org/abs/0801.4555}{{\ttfamily arXiv:0801.4555 [nucl-ex]}}.
%%CITATION = ARXIV:0801.4555;%%.

\bibitem{Adare:2012uk}
{\bfseries PHENIX} Collaboration, A.~Adare {\em et~al.}, ``{Evolution of
  $\pi^0$ suppression in Au+Au collisions from $\sqrt{s_{NN}} = 39$ to 200
  GeV},'' \href{http://dx.doi.org/10.1103/PhysRevLett.109.152301}{{\em
  Phys.Rev.Lett.} {\bfseries 109} (2012) 152301},
\href{http://arxiv.org/abs/1204.1526}{{\ttfamily arXiv:1204.1526 [nucl-ex]}}.
%%CITATION = ARXIV:1204.1526;%%.

\bibitem{Aamodt:2010jd}
{\bfseries ALICE} Collaboration, K.~Aamodt {\em et~al.}, ``{Suppression of
  Charged Particle Production at Large Transverse Momentum in Central Pb--Pb
  Collisions at $\sqrt{s_{NN}} = 2.76$ TeV},''
  \href{http://dx.doi.org/10.1016/j.physletb.2010.12.020}{{\em Phys.Lett.}
  {\bfseries B696} (2011) 30--39},
\href{http://arxiv.org/abs/1012.1004}{{\ttfamily arXiv:1012.1004 [nucl-ex]}}.
%%CITATION = ARXIV:1012.1004;%%.

\bibitem{CMS:2012aa}
{\bfseries CMS} Collaboration, S.~Chatrchyan {\em et~al.}, ``{Study of high-pT
  charged particle suppression in PbPb compared to $pp$ collisions at
  $\sqrt{s_{NN}}=2.76$ TeV},''
  \href{http://dx.doi.org/10.1140/epjc/s10052-012-1945-x}{{\em Eur.Phys.J.}
  {\bfseries C72} (2012) 1945},
\href{http://arxiv.org/abs/1202.2554}{{\ttfamily arXiv:1202.2554 [nucl-ex]}}.
%%CITATION = ARXIV:1202.2554;%%.

\bibitem{Abelev:2012hxa}
{\bfseries ALICE} Collaboration, B.~Abelev {\em et~al.}, ``{Centrality
  Dependence of Charged Particle Production at Large Transverse Momentum in
  Pb--Pb Collisions at $\sqrt{s_{\rm{NN}}} = 2.76$ TeV},''
  \href{http://dx.doi.org/10.1016/j.physletb.2013.01.051}{{\em Phys.Lett.}
  {\bfseries B720} (2013) 52--62},
\href{http://arxiv.org/abs/1208.2711}{{\ttfamily arXiv:1208.2711 [hep-ex]}}.
%%CITATION = ARXIV:1208.2711;%%.

\bibitem{Sharma:2009hn}
R.~Sharma, I.~Vitev, and B.-W. Zhang, ``{Light-cone wave function approach to
  open heavy flavor dynamics in QCD matter},''
  \href{http://dx.doi.org/10.1103/PhysRevC.80.054902}{{\em Phys.Rev.}
  {\bfseries C80} (2009) 054902},
\href{http://arxiv.org/abs/0904.0032}{{\ttfamily arXiv:0904.0032 [hep-ph]}}.
%%CITATION = ARXIV:0904.0032;%%.

\bibitem{Neufeld:2010dz}
R.~Neufeld, I.~Vitev, and B.-W. Zhang, ``{A possible determination of the quark
  radiation length in cold nuclear matter},''
  \href{http://dx.doi.org/10.1016/j.physletb.2011.09.045}{{\em Phys.Lett.}
  {\bfseries B704} (2011) 590--595},
\href{http://arxiv.org/abs/1010.3708}{{\ttfamily arXiv:1010.3708 [hep-ph]}}.
%%CITATION = ARXIV:1010.3708;%%.

\bibitem{Beringer:1900zz}
{\bfseries Particle Data Group} Collaboration, J.~Beringer {\em et~al.},
  ``{Review of Particle Physics (RPP)},''
\href{http://dx.doi.org/10.1103/PhysRevD.86.010001}{{\em Phys.Rev.} {\bfseries
  D86} (2012) 010001}.
%%CITATION = PHRVA,D86,010001;%%.

\bibitem{Dellacasa:1999kd}
{\bfseries ALICE} Collaboration, G.~Dellacasa {\em et~al.}, ``{ALICE technical
  design report of the photon spectrometer (PHOS)},''
{\em CERN-LHCC-99-04} (1999) .
%%CITATION = CERN-LHCC-99-04 ETC.;%%.

\bibitem{Aamodt:2010aa}
{\bfseries ALICE} Collaboration, K.~Aamodt {\em et~al.}, ``{Alignment of the
  ALICE Inner Tracking System with cosmic-ray tracks},''
  \href{http://dx.doi.org/10.1088/1748-0221/5/03/P03003}{{\em JINST} {\bfseries
  5} (2010) P03003},
\href{http://arxiv.org/abs/1001.0502}{{\ttfamily arXiv:1001.0502
  [physics.ins-det]}}.
%%CITATION = ARXIV:1001.0502;%%.

\bibitem{Alme:2010ke}
J.~Alme, Y.~Andres, H.~Appelshauser, S.~Bablok, N.~Bialas, {\em et~al.}, ``{The
  ALICE TPC, a large 3-dimensional tracking device with fast readout for
  ultra-high multiplicity events},''
  \href{http://dx.doi.org/10.1016/j.nima.2010.04.042}{{\em Nucl.Instrum.Meth.}
  {\bfseries A622} (2010) 316--367},
\href{http://arxiv.org/abs/1001.1950}{{\ttfamily arXiv:1001.1950
  [physics.ins-det]}}.
%%CITATION = ARXIV:1001.1950;%%.

\bibitem{Aamodt:2008zz}
{\bfseries ALICE} Collaboration, K.~Aamodt {\em et~al.}, ``{The ALICE
  experiment at the CERN LHC},''
\href{http://dx.doi.org/10.1088/1748-0221/3/08/S08002}{{\em JINST} {\bfseries
  3} (2008) S08002}.
%%CITATION = JINST,3,S08002;%%.

\bibitem{Cortese:2004aa}
{\bfseries ALICE} Collaboration, P.~Cortese {\em et~al.}, ``{ALICE technical
  design report on forward detectors: FMD, T0 and V0},''
{\em CERN-LHCC-2004-025} (2004) .
%%CITATION = CERN-LHCC-2004-025 ETC.;%%.

\bibitem{Abelev:2013qoq}
{\bfseries ALICE} Collaboration, B.~Abelev {\em et~al.}, ``{Centrality
  determination of Pb-Pb collisions at $\sqrt{s_{NN}}$ = 2.76 TeV with
  ALICE},'' \href{http://dx.doi.org/10.1103/PhysRevC.88.044909}{{\em Phys.Rev.}
  {\bfseries C88} (2013) 044909},
\href{http://arxiv.org/abs/1301.4361}{{\ttfamily arXiv:1301.4361 [nucl-ex]}}.
%%CITATION = ARXIV:1301.4361;%%.

\bibitem{Abelev:2012sea}
{\bfseries ALICE} Collaboration, B.~Abelev {\em et~al.}, ``{Measurement of
  inelastic, single- and double-diffraction cross sections in proton--proton
  collisions at the LHC with ALICE},''
  \href{http://dx.doi.org/10.1140/epjc/s10052-013-2456-0}{{\em Eur.Phys.J.}
  {\bfseries C73} (2013) 2456},
\href{http://arxiv.org/abs/1208.4968}{{\ttfamily arXiv:1208.4968 [hep-ex]}}.
%%CITATION = ARXIV:1208.4968;%%.

\bibitem{Abelev:2012cn}
{\bfseries ALICE} Collaboration, B.~Abelev {\em et~al.}, ``{Neutral pion and
  $\eta$ meson production in proton-proton collisions at $\sqrt{s}=0.9$ TeV and
  $\sqrt{s}=7$ TeV},''
  \href{http://dx.doi.org/10.1016/j.physletb.2012.09.015}{{\em Phys.Lett.}
  {\bfseries B717} (2012) 162--172},
\href{http://arxiv.org/abs/1205.5724}{{\ttfamily arXiv:1205.5724 [hep-ex]}}.
%%CITATION = ARXIV:1205.5724;%%.

\bibitem{Abelev:2014ffa}
{\bfseries ALICE} Collaboration, B.~B. Abelev {\em et~al.}, ``{Performance of
  the ALICE Experiment at the CERN LHC},''
\href{http://arxiv.org/abs/1402.4476}{{\ttfamily arXiv:1402.4476 [nucl-ex]}}.
%%CITATION = ARXIV:1402.4476;%%.

\bibitem{Brun:1987ma}
R.~Brun, F.~Bruyant, M.~Maire, A.~McPherson, and P.~Zanarini, ``{GEANT3},''
  Tech. Rep. CERN-DD-EE-84-1, CERN, 1987.

\bibitem{Alessandro:2006yt}
{\bfseries ALICE} Collaboration, E.~Alessandro, G {\em et~al.}, ``{ALICE:
  Physics performance report, volume II},''
\href{http://dx.doi.org/10.1088/0954-3899/32/10/001}{{\em J.Phys.G} {\bfseries
  G32} (2006) 1295--2040}.
%%CITATION = JPHGB,G32,1295;%%.

\bibitem{CBM2007}
S.~Gorbunov and I.~Kisel, ``{Reconstruction of decayed particles based on the
  Kalman filter},'' Tech. Rep. CBM-SOFT-note-2007-003, CBM experiment, 2007.

\bibitem{podolanski1954iii}
J.~Podolanski and R.~Armenteros, ``Iii. analysis of v-events,'' {\em
  Philosophical Magazine} {\bfseries 45} no.~360, (1954) 13--30.

\bibitem{CrystalBall:1980}
M.~J. Oreglia, {\em A Study of the Reactions $\psi\prime\to\gamma \gamma
  \psi$}.
\newblock PhD thesis, SLAC, Stanford University, Stanford, California 94305,
  1980.
\newblock \url{http://www.slac.stanford.edu/pubs/slacreports/slac-r-236.html}.

\bibitem{Koch:2011}
{\bfseries ALICE} Collaboration, K.~Koch, ``{$\pi^0$ and $\eta$ measurement
  with photon conversions in ALICE in proton-proton collisions at $\sqrt{s} =
  7$~TeV},'' \href{http://dx.doi.org/10.1016/j.nuclphysa.2011.02.059}{{\em
  Nucl.Phys.A} {\bfseries 855} (2011) 281--284}.

\bibitem{Sjostrand:2007gs}
T.~Sjostrand, S.~Mrenna, and P.~Z. Skands, ``{A Brief Introduction to PYTHIA
  8.1},'' \href{http://dx.doi.org/10.1016/j.cpc.2008.01.036}{{\em
  Comput.Phys.Commun.} {\bfseries 178} (2008) 852--867},
\href{http://arxiv.org/abs/0710.3820}{{\ttfamily arXiv:0710.3820 [hep-ph]}}.
%%CITATION = ARXIV:0710.3820;%%.

\bibitem{Engel:1995sb}
R.~Engel, J.~Ranft, and S.~Roesler, ``{Hard diffraction in hadron hadron
  interactions and in photoproduction},''
  \href{http://dx.doi.org/10.1103/PhysRevD.52.1459}{{\em Phys.Rev.} {\bfseries
  D52} (1995) 1459--1468},
\href{http://arxiv.org/abs/hep-ph/9502319}{{\ttfamily arXiv:hep-ph/9502319
  [hep-ph]}}.
%%CITATION = HEP-PH/9502319;%%.

\bibitem{Gyulassy:1994ew}
M.~Gyulassy and X.-N. Wang, ``{HIJING 1.0: A Monte Carlo program for parton and
  particle production in high-energy hadronic and nuclear collisions},''
  \href{http://dx.doi.org/10.1016/0010-4655(94)90057-4}{{\em
  Comput.Phys.Commun.} {\bfseries 83} (1994) 307},
\href{http://arxiv.org/abs/nucl-th/9502021}{{\ttfamily arXiv:nucl-th/9502021
  [nucl-th]}}.
%%CITATION = NUCL-TH/9502021;%%.

\bibitem{Lafferty:1994cj}
G.~Lafferty and T.~Wyatt, ``{Where to stick your data points: The treatment of
  measurements within wide bins},''
\href{http://dx.doi.org/10.1016/0168-9002(94)01112-5}{{\em Nucl.Instrum.Meth.}
  {\bfseries A355} (1995) 541--547}.
%%CITATION = NUIMA,A355,541;%%.

\bibitem{Abelev:2013xaa}
{\bfseries ALICE} Collaboration, B.~B. Abelev {\em et~al.}, ``{$K^0_S$ and
  $\Lambda$ production in Pb-Pb collisions at sqrt(sNN) = 2.76 TeV},''
  \href{http://dx.doi.org/10.1103/PhysRevLett.111.222301}{{\em Phys.Rev.Lett.}
  {\bfseries 111} (2013) 222301},
\href{http://arxiv.org/abs/1307.5530}{{\ttfamily arXiv:1307.5530 [nucl-ex]}}.
%%CITATION = ARXIV:1307.5530;%%.

\bibitem{d'Enterria:2013vba}
D.~d'Enterria, K.~J. Eskola, I.~Helenius, and H.~Paukkunen, ``{Confronting
  current NLO parton fragmentation functions with inclusive charged-particle
  spectra at hadron colliders},'' {\em Nucl.Phys.} {\bfseries B883} (2013) 615,
\href{http://arxiv.org/abs/1311.1415}{{\ttfamily arXiv:1311.1415 [hep-ph]}}.
%%CITATION = ARXIV:1311.1415;%%.

\bibitem{deFlorian:2007aj}
D.~de~Florian, R.~Sassot, and M.~Stratmann, ``{Global analysis of fragmentation
  functions for pions and kaons and their uncertainties},''
  \href{http://dx.doi.org/10.1103/PhysRevD.75.114010}{{\em Phys.Rev.}
  {\bfseries D75} (2007) 114010},
\href{http://arxiv.org/abs/hep-ph/0703242}{{\ttfamily arXiv:hep-ph/0703242
  [HEP-PH]}}.
%%CITATION = HEP-PH/0703242;%%.

\bibitem{Pumplin:2002vw}
J.~Pumplin, D.~Stump, J.~Huston, H.~Lai, P.~M. Nadolsky, {\em et~al.}, ``{New
  generation of parton distributions with uncertainties from global QCD
  analysis},'' \href{http://dx.doi.org/10.1088/1126-6708/2002/07/012}{{\em
  JHEP} {\bfseries 0207} (2002) 012},
\href{http://arxiv.org/abs/hep-ph/0201195}{{\ttfamily arXiv:hep-ph/0201195
  [hep-ph]}}.
%%CITATION = HEP-PH/0201195;%%.

\bibitem{Corke:2010yf}
R.~Corke and T.~Sjostrand, ``{Interleaved Parton Showers and Tuning
  Prospects},'' \href{http://dx.doi.org/10.1007/JHEP03(2011)032}{{\em JHEP}
  {\bfseries 1103} (2011) 032},
\href{http://arxiv.org/abs/1011.1759}{{\ttfamily arXiv:1011.1759 [hep-ph]}}.
%%CITATION = ARXIV:1011.1759;%%.

\bibitem{Lappi:2013zma}
T.~Lappi and H.~Mäntysaari, ``{Single inclusive particle production at high
  energy from HERA data to proton-nucleus collisions},''
  \href{http://dx.doi.org/10.1103/PhysRevD.88.114020}{{\em Phys.Rev.}
  {\bfseries D88} (2013) 114020},
\href{http://arxiv.org/abs/1309.6963}{{\ttfamily arXiv:1309.6963 [hep-ph]}}.
%%CITATION = ARXIV:1309.6963;%%.

\bibitem{Miller:2007ri}
M.~L. Miller, K.~Reygers, S.~J. Sanders, and P.~Steinberg, ``{Glauber modeling
  in high energy nuclear collisions},''
  \href{http://dx.doi.org/10.1146/annurev.nucl.57.090506.123020}{{\em
  Ann.Rev.Nucl.Part.Sci.} {\bfseries 57} (2007) 205--243},
\href{http://arxiv.org/abs/nucl-ex/0701025}{{\ttfamily arXiv:nucl-ex/0701025
  [nucl-ex]}}.
%%CITATION = NUCL-EX/0701025;%%.

\bibitem{Alver:2008aq}
B.~Alver, M.~Baker, C.~Loizides, and P.~Steinberg, ``{The PHOBOS Glauber Monte
  Carlo},''
\href{http://arxiv.org/abs/0805.4411}{{\ttfamily arXiv:0805.4411 [nucl-ex]}}.
%%CITATION = ARXIV:0805.4411;%%.

\bibitem{Abelev:2014laa}
{\bfseries ALICE} Collaboration, B.~B. Abelev {\em et~al.}, ``{Production of
  charged pions, kaons and protons at large transverse momenta in pp and Pb-Pb
  collisions at sqrt(sNN) = 2.76 TeV},''
\href{http://arxiv.org/abs/1401.1250}{{\ttfamily arXiv:1401.1250 [nucl-ex]}}.
%%CITATION = ARXIV:1401.1250;%%.

\bibitem{Aggarwal:2007gw}
{\bfseries WA98} Collaboration, M.~Aggarwal {\em et~al.}, ``{Suppression of
  High-p(T) Neutral Pions in Central Pb+Pb Collisions at s(NN)**(1/2) =
  17.3-GeV},'' \href{http://dx.doi.org/10.1103/PhysRevLett.100.242301}{{\em
  Phys.Rev.Lett.} {\bfseries 100} (2008) 242301},
\href{http://arxiv.org/abs/0708.2630}{{\ttfamily arXiv:0708.2630 [nucl-ex]}}.
%%CITATION = ARXIV:0708.2630;%%.

\bibitem{Adler:2006bw}
{\bfseries PHENIX} Collaboration, S.~Adler {\em et~al.}, ``{A Detailed Study of
  High-p(T) Neutral Pion Suppression and Azimuthal Anisotropy in Au+Au
  Collisions at s(NN)**(1/2) = 200-GeV},''
  \href{http://dx.doi.org/10.1103/PhysRevC.76.034904}{{\em Phys.Rev.}
  {\bfseries C76} (2007) 034904},
\href{http://arxiv.org/abs/nucl-ex/0611007}{{\ttfamily arXiv:nucl-ex/0611007
  [nucl-ex]}}.
%%CITATION = NUCL-EX/0611007;%%.

\bibitem{Horowitz:2007nq}
W.~A. Horowitz, ``{LHC Predictions from an extended theory with Elastic,
  Inelastic, and Path Length Fluctuating Energy Loss},''
  \href{http://dx.doi.org/10.1142/S0218301307007672}{{\em Int.J.Mod.Phys.}
  {\bfseries E16} (2007) 2193--2199},
\href{http://arxiv.org/abs/nucl-th/0702084}{{\ttfamily arXiv:nucl-th/0702084
  [NUCL-TH]}}.
%%CITATION = NUCL-TH/0702084;%%.

\bibitem{Werner:2012xh}
K.~Werner, I.~Karpenko, M.~Bleicher, T.~Pierog, and S.~Porteboeuf-Houssais,
  ``{Jets, Bulk Matter, and their Interaction in Heavy Ion Collisions at
  Several TeV},'' \href{http://dx.doi.org/10.1103/PhysRevC.85.064907}{{\em
  Phys.Rev.} {\bfseries C85} (2012) 064907},
\href{http://arxiv.org/abs/1203.5704}{{\ttfamily arXiv:1203.5704 [nucl-th]}}.
%%CITATION = ARXIV:1203.5704;%%.

\bibitem{Kopeliovich:2012sc}
B.~Kopeliovich, J.~Nemchik, I.~Potashnikova, and I.~Schmidt, ``{Quenching of
  high-pT hadrons: Energy Loss vs Color Transparency},''
  \href{http://dx.doi.org/10.1103/PhysRevC.86.054904}{{\em Phys.Rev.}
  {\bfseries C86} (2012) 054904},
\href{http://arxiv.org/abs/1208.4951}{{\ttfamily arXiv:1208.4951 [hep-ph]}}.
%%CITATION = ARXIV:1208.4951;%%.

\bibitem{Nemchik:2013ooa}
J.~Nemchik, I.~A. Karpenko, B.~Kopeliovich, I.~Potashnikova, and Y.~M.
  Sinyukov, ``{High-pT hadrons from nuclear collisions: Unifying pQCD with
  hydrodynamics},''
\href{http://arxiv.org/abs/1310.3455}{{\ttfamily arXiv:1310.3455 [hep-ph]}}.
%%CITATION = ARXIV:1310.3455;%%.

\bibitem{Abelev:2013ala}
{\bfseries ALICE} Collaboration, B.~B. Abelev {\em et~al.}, ``{Energy
  Dependence of the Transverse Momentum Distributions of Charged Particles in
  pp Collisions Measured by ALICE},''
  \href{http://dx.doi.org/10.1140/epjc/s10052-013-2662-9}{{\em Eur.Phys.J.}
  {\bfseries C73} (2013) 2662},
\href{http://arxiv.org/abs/1307.1093}{{\ttfamily arXiv:1307.1093 [nucl-ex]}}.
%%CITATION = ARXIV:1307.1093;%%.

\end{thebibliography}\endgroup

\onecolumn

\appendix
\section{Extrapolation}
For the calculation of the $\RAA$ above $\pT > \unit[8]{GeV}/c$ an extrapolation
of the measured transverse momentum spectrum in pp collisions at $\s = \unit[2.76]{TeV}$ 
based on the Tsallis functional form  
\begin{eqnarray}
  \frac{1}{2 \pi \pT} \frac{\mathrm{d}^2N}{\mathrm{d} \pT \mathrm{d}y} &=&
  \frac{A}{2\pi} 
   \frac{(n-1)(n-2)}{nC\left[ nC+m(n-2)\right]}  \frac{1}{c^2}\nonumber \\ 
    && \cdot \left(1+\frac{\sqrt{\pT^2+m^2}-m}{nC}\right)^{-n}
    \label{eq:Tsallis}
\end{eqnarray}
was used (where $m$ is the mass of the neutral pion and $c$ the speed of light). 
The parameters are given in Table~\ref{tab:TsallisParam}.

\begin{table}[t]
  \centering
  \begin{tabular}{|l|r|r|r|} \hline
    system & $A$  & $C$ (MeV/$c^2$)  & $n$ \\ \hline
    pp & $1.7 \pm 0.7$   & $ 135 \pm 29$ & $7.1 \pm 0.7$ \\\hline
    $60-80\%$ \PbPb & $31.7$ &	$142$ & $7.4$ 	 \\\hline
  \end{tabular}
  \caption{Parameters of the fits of the  Tsallis parameterization
    (Eq.~\ref{eq:Tsallis}) to the combined invariant production yields for
    $\pi^0$ mesons in inelastic collisions at $\s = \unit[2.76]{TeV}$. 
    The uncertainties (statistical and systematic added in quadrature) were
    used to evaluate the uncertainty of the extrapolation used in the calculation
    of $\RAA$ for $\pT > \unit[8]{GeV}/c$. The uncertainty on the parameter $A$
    due to the spectra normalization 
    of $3.9\%$ at $\s = \unit[2.76]{TeV}$ is not included. For the measurment in 
    $60-80\%$~\PbPb~collisions the fit parameters are given without uncertainties 
    as the parameterization is only used to facilitate the comparison with model calculations.}
  \label{tab:TsallisParam}
\end{table}
\begin{table}[t]
  \centering
  \begin{tabular}{|r|r|r|r|r|r|} \hline
    centrality & $a$ ($c^2/\mathrm{GeV}^2$)      & $b$  & $c$ & $d$ & $e$\\ \hline
    $0-5\%$ & $28.96$ &	$5.85$ &	$-199.17$ &	$4.64$ &	$95.30$ 	\\ \hline
    $5-10\%$ & $21.97$ &	$5.79$ &	$-33.54$ &	$2.96$ &	$10.84$ 	\\ \hline
    $0-10\%$ & $25.53$ &	$5.84$ &	$-49.95$ &	$3.35$ &	$18.49$	\\ \hline
    $10-20\%$ & $18.91$ &	$5.71$ &	$-44.76$ &	$3.37$ &	$19.66$ 	\\ \hline
    $20-40\%$ & $11.54$ &	$5.74$ &	$-18.43$ &	$2.62$ &	$7.37$ 	\\ \hline
    $40-60\%$ & $4.18$ &	$5.67$ &	$-9.43$ &	$2.00$ &	$3.39$ 	\\ \hline
  \end{tabular}
  \caption{Parameters of the fits to the combined invariant yields of
    $\pi^0$ mesons in \PbPb\ collisions in different centrality classes with the
    functional form given in Eq.~\ref{eq:qcd}. The spectra were fitted taking into account 
    the combined statistical and systematic errors. }
  \label{tab:QCDParameters}
\end{table}

In order to compare the individual PCM and PHOS  measurements to the combined results in
\PbPb~collisions the parameterization 
\begin{equation}
  \frac{1}{2 \pi \pT} \frac{\mathrm{d}^2N}{\mathrm{d} \pT \mathrm{d}y}  = a \cdot \pT^{-(b+c/(\pT^d+e))}
 \label{eq:qcd}
\end{equation}
with $\pT$ in GeV/$c$ was used to fit the combined spectrum for each centrality class. 
The corresponding parameters are given in Tab.~\ref{tab:QCDParameters}. 
For the most peripheral centrality class the Tsallis 
parameterization Eq.~\ref{eq:Tsallis} was used for which the parameters are given in Tab.~\ref{tab:TsallisParam}. 
These parameterizations describe the data well in the measured momentum range.

\newpage
\section{The ALICE Collaboration}
\label{app:collab}

% Collaboration: CERN-LHC-ALICE
% Generation Date is 2014/Apr/28

% How to use:
%%%%%%%%% appendix with author list
%\appendix
%\section{The ALICE Collaboration}
%\label{app:collab}
%\input{authors-list.tex}  %%%%%%% get the latest version before submitting

\begingroup
\small
\begin{flushleft}
B.~Abelev\Irefn{org69}\And
J.~Adam\Irefn{org37}\And
D.~Adamov\'{a}\Irefn{org77}\And
M.M.~Aggarwal\Irefn{org81}\And
M.~Agnello\Irefn{org105}\textsuperscript{,}\Irefn{org88}\And
A.~Agostinelli\Irefn{org26}\And
N.~Agrawal\Irefn{org44}\And
Z.~Ahammed\Irefn{org124}\And
N.~Ahmad\Irefn{org18}\And
I.~Ahmed\Irefn{org15}
% ...
\renewcommand\labelenumi{\textsuperscript{\theenumi}~}

\section*{Affiliation notes}
\renewcommand\theenumi{\roman{enumi}}
\begin{Authlist}
\item \Adef{0}Deceased
\item \Adef{idp1098208}{Also at: St. Petersburg State Polytechnical University}
\item \Adef{idp2995200}{Also at: Department of Applied Physics, Aligarh Muslim University, Aligarh, India}
\item \Adef{idp3684128}{Also at: M.V. Lomonosov Moscow State University, D.V. Skobeltsyn Institute of Nuclear Physics, Moscow, Russia}
\item \Adef{idp3927728}{Also at: University of Belgrade, Faculty of Physics and "Vin\v{c}a" Institute of Nuclear Sciences, Belgrade, Serbia}
\item \Adef{idp4233280}{Permanent Address: Permanent Address: Konkuk University, Seoul, Korea}
\item \Adef{idp4773664}{Also at: Institute of Theoretical Physics, University of Wroclaw, Wroclaw, Poland}
\item \Adef{idp5691024}{Also at: University of Kansas, Lawrence, KS, United States}
\end{Authlist}

\section*{Collaboration Institutes}
\renewcommand\theenumi{\arabic{enumi}~}
\begin{Authlist}

\item \Idef{org1}A.I. Alikhanyan National Science Laboratory (Yerevan Physics Institute) Foundation, Yerevan, Armenia
\item \Idef{org2}Benem\'{e}rita Universidad Aut\'{o}noma de Puebla, Puebla, Mexico
\item \Idef{org3}Bogolyubov Institute for Theoretical Physics, Kiev, Ukraine
\item \Idef{org4}Bose Institute, Department of Physics and Centre for Astroparticle Physics and Space Science (CAPSS), Kolkata, India
\item \Idef{org5}Budker Institute for Nuclear Physics, Novosibirsk, Russia
\item \Idef{org6}California Polytechnic State University, San Luis Obispo, CA, United States
\item \Idef{org7}Central China Normal University, Wuhan, China
\item \Idef{org8}Centre de Calcul de l'IN2P3, Villeurbanne, France
\item \Idef{org9}Centro de Aplicaciones Tecnol\'{o}gicas y Desarrollo Nuclear (CEADEN), Havana, Cuba
\item \Idef{org10}Centro de Investigaciones Energ\'{e}ticas Medioambientales y Tecnol\'{o}gicas (CIEMAT), Madrid, Spain
\item \Idef{org11}Centro de Investigaci\'{o}n y de Estudios Avanzados (CINVESTAV), Mexico City and M\'{e}rida, Mexico
\item \Idef{org12}Centro Fermi - Museo Storico della Fisica e Centro Studi e Ricerche ``Enrico Fermi'', Rome, Italy
\item \Idef{org13}Chicago State University, Chicago, USA
\item \Idef{org14}Commissariat \`{a} l'Energie Atomique, IRFU, Saclay, France
\item \Idef{org15}COMSATS Institute of Information Technology (CIIT), Islamabad, Pakistan
\item \Idef{org16}Departamento de F\'{\i}sica de Part\'{\i}culas and IGFAE, Universidad de Santiago de Compostela, Santiago de Compostela, Spain
\item \Idef{org17}Department of Physics and Technology, University of Bergen, Bergen, Norway
\item \Idef{org18}Department of Physics, Aligarh Muslim University, Aligarh, India
\item \Idef{org19}Department of Physics, Ohio State University, Columbus, OH, United States
\item \Idef{org20}Department of Physics, Sejong University, Seoul, South Korea
\item \Idef{org21}Department of Physics, University of Oslo, Oslo, Norway
\item \Idef{org22}Dipartimento di Fisica dell'Universit\`{a} 'La Sapienza' and Sezione INFN Rome, Italy
\item \Idef{org23}Dipartimento di Fisica dell'Universit\`{a} and Sezione INFN, Cagliari, Italy
\item \Idef{org24}Dipartimento di Fisica dell'Universit\`{a} and Sezione INFN, Trieste, Italy
\item \Idef{org25}Dipartimento di Fisica dell'Universit\`{a} and Sezione INFN, Turin, Italy
\item \Idef{org26}Dipartimento di Fisica e Astronomia dell'Universit\`{a} and Sezione INFN, Bologna, Italy
\item \Idef{org27}Dipartimento di Fisica e Astronomia dell'Universit\`{a} and Sezione INFN, Catania, Italy
\item \Idef{org28}Dipartimento di Fisica e Astronomia dell'Universit\`{a} and Sezione INFN, Padova, Italy
\item \Idef{org29}Dipartimento di Fisica `E.R.~Caianiello' dell'Universit\`{a} and Gruppo Collegato INFN, Salerno, Italy
\item \Idef{org30}Dipartimento di Scienze e Innovazione Tecnologica dell'Universit\`{a} del  Piemonte Orientale and Gruppo Collegato INFN, Alessandria, Italy
\item \Idef{org31}Dipartimento Interateneo di Fisica `M.~Merlin' and Sezione INFN, Bari, Italy
\item \Idef{org32}Division of Experimental High Energy Physics, University of Lund, Lund, Sweden
\item \Idef{org33}Eberhard Karls Universit\"{a}t T\"{u}bingen, T\"{u}bingen, Germany
\item \Idef{org34}European Organization for Nuclear Research (CERN), Geneva, Switzerland
\item \Idef{org35}Faculty of Engineering, Bergen University College, Bergen, Norway
\item \Idef{org36}Faculty of Mathematics, Physics and Informatics, Comenius University, Bratislava, Slovakia
\item \Idef{org37}Faculty of Nuclear Sciences and Physical Engineering, Czech Technical University in Prague, Prague, Czech Republic
\item \Idef{org38}Faculty of Science, P.J.~\v{S}af\'{a}rik University, Ko\v{s}ice, Slovakia
\item \Idef{org39}Frankfurt Institute for Advanced Studies, Johann Wolfgang Goethe-Universit\"{a}t Frankfurt, Frankfurt, Germany
\item \Idef{org40}Gangneung-Wonju National University, Gangneung, South Korea
\item \Idef{org41}Gauhati University, Department of Physics, Guwahati, India
\item \Idef{org42}Helsinki Institute of Physics (HIP), Helsinki, Finland
\item \Idef{org43}Hiroshima University, Hiroshima, Japan
\item \Idef{org44}Indian Institute of Technology Bombay (IIT), Mumbai, India
\item \Idef{org45}Indian Institute of Technology Indore, Indore (IITI), India
\item \Idef{org46}Institut de Physique Nucl\'eaire d'Orsay (IPNO), Universit\'e Paris-Sud, CNRS-IN2P3, Orsay, France
\item \Idef{org47}Institut f\"{u}r Informatik, Johann Wolfgang Goethe-Universit\"{a}t Frankfurt, Frankfurt, Germany
\item \Idef{org48}Institut f\"{u}r Kernphysik, Johann Wolfgang Goethe-Universit\"{a}t Frankfurt, Frankfurt, Germany
\item \Idef{org49}Institut f\"{u}r Kernphysik, Westf\"{a}lische Wilhelms-Universit\"{a}t M\"{u}nster, M\"{u}nster, Germany
\item \Idef{org50}Institut Pluridisciplinaire Hubert Curien (IPHC), Universit\'{e} de Strasbourg, CNRS-IN2P3, Strasbourg, France
\item \Idef{org51}Institute for Nuclear Research, Academy of Sciences, Moscow, Russia
\item \Idef{org52}Institute for Subatomic Physics of Utrecht University, Utrecht, Netherlands
\item \Idef{org53}Institute for Theoretical and Experimental Physics, Moscow, Russia
\item \Idef{org54}Institute of Experimental Physics, Slovak Academy of Sciences, Ko\v{s}ice, Slovakia
\item \Idef{org55}Institute of Physics, Academy of Sciences of the Czech Republic, Prague, Czech Republic
\item \Idef{org56}Institute of Physics, Bhubaneswar, India
\item \Idef{org57}Institute of Space Science (ISS), Bucharest, Romania
\item \Idef{org58}Instituto de Ciencias Nucleares, Universidad Nacional Aut\'{o}noma de M\'{e}xico, Mexico City, Mexico
\item \Idef{org59}Instituto de F\'{\i}sica, Universidad Nacional Aut\'{o}noma de M\'{e}xico, Mexico City, Mexico
\item \Idef{org60}iThemba LABS, National Research Foundation, Somerset West, South Africa
\item \Idef{org61}Joint Institute for Nuclear Research (JINR), Dubna, Russia
\item \Idef{org62}Korea Institute of Science and Technology Information, Daejeon, South Korea
\item \Idef{org63}KTO Karatay University, Konya, Turkey
\item \Idef{org64}Laboratoire de Physique Corpusculaire (LPC), Clermont Universit\'{e}, Universit\'{e} Blaise Pascal, CNRS--IN2P3, Clermont-Ferrand, France
\item \Idef{org65}Laboratoire de Physique Subatomique et de Cosmologie, Universit\'{e} Grenoble-Alpes, CNRS-IN2P3, Grenoble, France
\item \Idef{org66}Laboratori Nazionali di Frascati, INFN, Frascati, Italy
\item \Idef{org67}Laboratori Nazionali di Legnaro, INFN, Legnaro, Italy
\item \Idef{org68}Lawrence Berkeley National Laboratory, Berkeley, CA, United States
\item \Idef{org69}Lawrence Livermore National Laboratory, Livermore, CA, United States
\item \Idef{org70}Moscow Engineering Physics Institute, Moscow, Russia
\item \Idef{org71}National Centre for Nuclear Studies, Warsaw, Poland
\item \Idef{org72}National Institute for Physics and Nuclear Engineering, Bucharest, Romania
\item \Idef{org73}National Institute of Science Education and Research, Bhubaneswar, India
\item \Idef{org74}Niels Bohr Institute, University of Copenhagen, Copenhagen, Denmark
\item \Idef{org75}Nikhef, National Institute for Subatomic Physics, Amsterdam, Netherlands
\item \Idef{org76}Nuclear Physics Group, STFC Daresbury Laboratory, Daresbury, United Kingdom
\item \Idef{org77}Nuclear Physics Institute, Academy of Sciences of the Czech Republic, \v{R}e\v{z} u Prahy, Czech Republic
\item \Idef{org78}Oak Ridge National Laboratory, Oak Ridge, TN, United States
\item \Idef{org79}Petersburg Nuclear Physics Institute, Gatchina, Russia
\item \Idef{org80}Physics Department, Creighton University, Omaha, NE, United States
\item \Idef{org81}Physics Department, Panjab University, Chandigarh, India
\item \Idef{org82}Physics Department, University of Athens, Athens, Greece
\item \Idef{org83}Physics Department, University of Cape Town, Cape Town, South Africa
\item \Idef{org84}Physics Department, University of Jammu, Jammu, India
\item \Idef{org85}Physics Department, University of Rajasthan, Jaipur, India
\item \Idef{org86}Physik Department, Technische Universit\"{a}t M\"{u}nchen, Munich, Germany
\item \Idef{org87}Physikalisches Institut, Ruprecht-Karls-Universit\"{a}t Heidelberg, Heidelberg, Germany
\item \Idef{org88}Politecnico di Torino, Turin, Italy
\item \Idef{org89}Purdue University, West Lafayette, IN, United States
\item \Idef{org90}Pusan National University, Pusan, South Korea
\item \Idef{org91}Research Division and ExtreMe Matter Institute EMMI, GSI Helmholtzzentrum f\"ur Schwerionenforschung, Darmstadt, Germany
\item \Idef{org92}Rudjer Bo\v{s}kovi\'{c} Institute, Zagreb, Croatia
\item \Idef{org93}Russian Federal Nuclear Center (VNIIEF), Sarov, Russia
\item \Idef{org94}Russian Research Centre Kurchatov Institute, Moscow, Russia
\item \Idef{org95}Saha Institute of Nuclear Physics, Kolkata, India
\item \Idef{org96}School of Physics and Astronomy, University of Birmingham, Birmingham, United Kingdom
\item \Idef{org97}Secci\'{o}n F\'{\i}sica, Departamento de Ciencias, Pontificia Universidad Cat\'{o}lica del Per\'{u}, Lima, Peru
\item \Idef{org98}Sezione INFN, Bari, Italy
\item \Idef{org99}Sezione INFN, Bologna, Italy
\item \Idef{org100}Sezione INFN, Cagliari, Italy
\item \Idef{org101}Sezione INFN, Catania, Italy
\item \Idef{org102}Sezione INFN, Padova, Italy
\item \Idef{org103}Sezione INFN, Rome, Italy
\item \Idef{org104}Sezione INFN, Trieste, Italy
\item \Idef{org105}Sezione INFN, Turin, Italy
\item \Idef{org106}SSC IHEP of NRC Kurchatov institute, Protvino, Russia
\item \Idef{org107}SUBATECH, Ecole des Mines de Nantes, Universit\'{e} de Nantes, CNRS-IN2P3, Nantes, France
\item \Idef{org108}Suranaree University of Technology, Nakhon Ratchasima, Thailand
\item \Idef{org109}Technical University of Split FESB, Split, Croatia
\item \Idef{org110}The Henryk Niewodniczanski Institute of Nuclear Physics, Polish Academy of Sciences, Cracow, Poland
\item \Idef{org111}The University of Texas at Austin, Physics Department, Austin, TX, USA
\item \Idef{org112}Universidad Aut\'{o}noma de Sinaloa, Culiac\'{a}n, Mexico
\item \Idef{org113}Universidade de S\~{a}o Paulo (USP), S\~{a}o Paulo, Brazil
\item \Idef{org114}Universidade Estadual de Campinas (UNICAMP), Campinas, Brazil
\item \Idef{org115}University of Houston, Houston, TX, United States
\item \Idef{org116}University of Jyv\"{a}skyl\"{a}, Jyv\"{a}skyl\"{a}, Finland
\item \Idef{org117}University of Liverpool, Liverpool, United Kingdom
\item \Idef{org118}University of Tennessee, Knoxville, TN, United States
\item \Idef{org119}University of Tokyo, Tokyo, Japan
\item \Idef{org120}University of Tsukuba, Tsukuba, Japan
\item \Idef{org121}University of Zagreb, Zagreb, Croatia
\item \Idef{org122}Universit\'{e} de Lyon, Universit\'{e} Lyon 1, CNRS/IN2P3, IPN-Lyon, Villeurbanne, France
\item \Idef{org123}V.~Fock Institute for Physics, St. Petersburg State University, St. Petersburg, Russia
\item \Idef{org124}Variable Energy Cyclotron Centre, Kolkata, India
\item \Idef{org125}Vestfold University College, Tonsberg, Norway
\item \Idef{org126}Warsaw University of Technology, Warsaw, Poland
\item \Idef{org127}Wayne State University, Detroit, MI, United States
\item \Idef{org128}Wigner Research Centre for Physics, Hungarian Academy of Sciences, Budapest, Hungary
\item \Idef{org129}Yale University, New Haven, CT, United States
\item \Idef{org130}Yonsei University, Seoul, South Korea
\item \Idef{org131}Zentrum f\"{u}r Technologietransfer und Telekommunikation (ZTT), Fachhochschule Worms, Worms, Germany
\end{Authlist}
\endgroup

  %%%%%%% get the latest version before submitting

% \tableofcontents

\end{document}